\documentclass[twocolumn,aps,prd,showpacs,epsfig,nofootinbib]{revtex4-1}
\usepackage{graphicx,epstopdf,latexsym,amssymb,amsmath,color,mathrsfs,rotating,multirow}
\usepackage[center]{subfigure}
 \usepackage[colorlinks,linkcolor=blue,citecolor=blue,urlcolor=blue]{hyperref}

\begin{document}
\allowdisplaybreaks
 \newcommand{\bq}{\begin{equation}}
 \newcommand{\eq}{\end{equation}}
 \newcommand{\bqn}{\begin{eqnarray}}
 \newcommand{\eqn}{\end{eqnarray}}
 \newcommand{\nb}{\nonumber}
 \newcommand{\lb}{\label}
 \newcommand{\f}{\frac}
 \newcommand{\p}{\partial}
\newcommand{\PRL}{Phys. Rev. Lett.}
\newcommand{\PLB}{Phys. Lett. B}
\newcommand{\PRD}{Phys. Rev. D}
\newcommand{\CQG}{Class. Quantum Grav.}
\newcommand{\JCAP}{J. Cosmol. Astropart. Phys.}
\newcommand{\JHEP}{J. High. Energy. Phys.}

\title{Ho\v{r}ava Gravity at a Lifshitz Point:  A Progress Report} 
\author{Anzhong Wang}   
\email{Anzhong$\_$Wang@baylor.edu} 
\affiliation{Institute  for Advanced Physics $\&$ Mathematics, Zhejiang University of Technology, Hangzhou 310032,  China\\ 
GCAP-CASPER, Department of Physics, Baylor University,
Waco, Texas 76798-7316, USA }
 
\date{\today}

\begin{abstract}

Ho\v{r}ava gravity at a Lifshitz  point  is a theory intended to quantize gravity by using techniques of  traditional quantum field theories. To
avoid Ostrogradsky's ghosts, a problem that has been plaguing quantization of general relativity since the middle of 1970's, Ho\v{r}ava chose to 
break the Lorentz invariance by a Lifshitz-type of anisotropic  scaling between space and time at the ultra-high energy, while recovering  
(approximately)  the invariance  at low energies. With the  stringent observational constraints and self-consistency, it turns out that this  is not an 
easy task, and various modifications have been proposed, since the  first incarnation of the theory in 2009.  In this  review,  we shall provide a 
progress report  on  the recent developments of Ho\v{r}ava gravity.  In particular, we first present four  most-studied versions of Ho\v{r}ava 
gravity, by focusing first on their self-consistency and then their consistency with experiments, including the solar system tests and cosmological
observations. Then, we provide a general review on the recent developments of the theory  in three different but also related  areas: (i) universal 
horizons, black holes and their  thermodynamics; (ii) non-relativistic gauge/gravity duality; and (iii) quantization of the theory. The studies in these 
areas can be generalized to other gravitational theories with broken Lorentz invariance.

\end{abstract}

\maketitle

 \tableofcontents

\section{Introduction}
\renewcommand{\theequation}{1.\arabic{equation}} \setcounter{equation}{0}

In the beginning  of the last century, physics started with two triumphs, {\em quantum mechanics and  general relativity}. On one hand, based on quantum mechanics
(QM),  the Standard Model (SM) of Particle Physics was developed, which describes  three of the four interactions: electromagnetism, and the weak and strong nuclear forces.
The last particle predicted by SM,  the Higgs boson, was  finally observed by  Large Hadron Collider in 2012 \cite{LHCa,LHCb}, after 40 years search. On the other hand, 
general relativity (GR)   describes the fourth force, gravity,  and predicts the existence of cosmic microwave background radiation (CMB), black holes, and gravitational waves
(GWs), among other things.  CMB was first observed  accidentally in 1964 \cite{PW}, and since then various experiments have remeasured it each time  
with unprecedented precisions \cite{cosmoAa,cosmoAb,cosmoBa,cosmoBb,cosmoC}. 
Black holes have attracted a great deal of attention both theoretically  and experimentally \cite{BHs1}, and various evidences of their existence were found \cite{BHs2}.
In particular, on Sept. 14,  2015,  the LIGO gravitational wave observatory made the first-ever successful observation of GWs \cite{GW}. The signal 
was consistent with theoretical predictions for the GWs produced by the merger of two binary black holes, which marks the beginning of a new era: {\em gravitational wave astronomy}.

Despite these spectacular  successes, we  have also been facing serious challenges. First, observations found that our universe consists of about $25\%$ 
dark matter (DM) \cite{cosmoC}. It is generally believed  that such matter  should not be made of particles from SM with a very simple argument: otherwise
 we should have already observed them directly.  
Second,   spacetime singularities exist generically  \cite{HE73},  including those of  black hole and the big bang cosmology. At the singularities, GR as well as 
any of other  physics laws are all broken down, and it has been a cherished hope    that quantum gravitational effects will step in and resolve the singularity problem.
 
However,  when applying the well-understood quantum field theories (QFTs)  to GR to obtain  a theory of quantum gravity (QG),  we have been facing  a tremendous resistance  
\cite{QGa,QGb,QGc}: GR is not (perturbatively) renormalizable. Power-counting analysis shows that this happens because in four-dimensional  spacetimes
 the gravitational coupling constant  $G_N$ has the dimension of ${\mbox{(mass)}}^{-2}$ (in units where the Planck constant $\hbar$ and the speed of light $c$  are one),
 whereas it should be larger than or equal to zero in order  for the theory to be renormalizable perturbatively  \cite{AS}. In fact, 
 the expansion of a given physical quantity $F$ in terms of the small gravitational coupling constant $G_N$ must be in the form 
 \bq
 \lb{1.1}
 F = \sum_{n=0}^{\infty}{a_n \left(G_NE^2\right)^{n}}, 
 \eq
 where $E$ denotes the energy of the  system involved, so that the combination $\left(G_NE^2\right)$ is dimensionless. Clearly, when $E^2 \gtrsim G_N^{-1}$, such  
 expansions  diverge. Therefore, it is expected that   perturbative effective QFT is broken  down at such energies.  It is  in this sense that  GR is often said 
  not perturbatively renormalizable.  
 
 An improved ultraviolet (UV) behavior can be obtained by including high-order derivative corrections to the Einstein-Hilbert action, 
 \bq
 \lb{1.2}
 S_{EH} = \int{d^4x \sqrt{-g} R},
 \eq
 such as a quadratic term,  $R_{\mu\nu}R^{\mu\nu}$ \cite{Stelle}. Then, the gravitational propagator will be changed from $1/k^{2}$ to 
  \bqn
 \lb{1.3}
 && \frac{1}{k^2} + \frac{1}{k^2} G_Nk^4\frac{1}{k^2} + \frac{1}{k^2} G_Nk^4\frac{1}{k^2}G_Nk^4\frac{1}{k^2} + ....\nb\\
 && ~~~~~~~~~~~~~~~~~~~~~~ = \frac{1}{k^2 - G_Nk^4}.  
 \eqn
Thus, at high energy the propagator is dominated by the term $1/k^4$, and as a result, the UV divergence can be cured. Unfortunately, this simultaneously makes the modified
theory not unitary, as now we have two poles, 
  \bqn
 \lb{1.4}
  \frac{1}{k^2 - G_Nk^4} =  \frac{1}{k^2} -  \frac{1}{k^2 - G_N^{-1}},
 \eqn
and the first one ($1/k^2$) describes a massless spin-2 graviton, while the second one describes a massive one but with a wrong sign in front of it, which implies that the 
massive graviton is actually a ghost (with a negative kinetic energy). It is the existence of this ghost that makes the theory not unitary, and has been  there since its discovery
\cite{Stelle}.  

The existence of the ghost is closely  related to the fact that the modified theory has orders of time-derivatives higher than two. In the quadratic case, for example,  the field equations  are  
 fourth-orders. As a matter of fact, there exists a powerful theorem due to Mikhail Vasilevich Ostrogradsky,  who established it   in 1850  \cite{Ostrogradsky}. 
 The theorem basically states that {\em a system is not (kinematically) stable if it is described by a non-degenerate higher time-derivative Lagrangian}. To be
more specific, let us consider a system whose Lagrangian depends on $\ddot{x}$, i.e., ${\cal{L}} = {\cal{L}}(x, \dot{x}, \ddot{x})$, where $\dot{x} = dx(t)/dt$, etc. Then, the Euler-Lagrange
equation reads,
\bq
\lb{1.5}
\frac{\partial{\cal{L}}}{\partial x} - \frac{d}{dt} \frac{\partial{\cal{L}}}{\partial \dot{x}} +  \frac{d^2}{dt^2} \frac{\partial{\cal{L}}}{\partial \ddot{x}} = 0.
\eq
Now the non-degeneracy means that ${\partial^2{\cal{L}}}/{\partial \ddot{x}^2} \not= 0$, which implies that Eq.(\ref{1.5}) can be cast in the form  \cite{Woodard15},
\bq
\lb{1.6}
\frac{d^4 x(t)}{dt^4} = {\cal{F}}(x, \dot{x}, \ddot{x}, \dddot{x}; t).
\eq
Clearly, in order to determine a solution uniquely four initial conditions are needed. This in turn implies that there
must be four canonical coordinates, which can be chosen as,
\bqn
\lb{1.7}
X_1 &\equiv& x, \quad P_1 \equiv   \frac{\partial{\cal{L}}}{\partial \dot{x}} -  \frac{d}{dt} \frac{\partial{\cal{L}}}{\partial \ddot{x}}, \nb\\
X_2 &\equiv& \dot{x}, \quad P_2 \equiv    \frac{\partial{\cal{L}}}{\partial \ddot{x}}.
\eqn
The assumption of non-degeneracy guarantees that Eq.(\ref{1.7}) has the inverse solution $\ddot{x} = A(X_1, X_2, P_2)$, so that 
\bqn
\lb{1.8}
\left.\frac{\partial{\cal{L}}}{\partial \ddot{x}}\right|_{x = X_1, \; \dot{x} = X_2,\; \ddot{x} =A} = P_2.
 \eqn
Then, the corresponding Hamiltonian is given by
\bqn
\lb{1.9}
H = \sum_{i=1}^{2}{P_i \frac{d^ix}{dt^i}} - {\cal{L}} = P_1X_2 + P_2A - {\cal{L}},
\eqn
which is linear in the canonical momentum $P_1$ and implies that  there are no barriers to prevent the system from decay, so the system is not stable
generically. 

It is remarkable to note how  powerful and  general that the theorem is:  {\em It
applies to any Lagrangian of the form ${\cal{L}}(x, \dot{x}, \ddot{x})$. The only    assumption is the non-degeneracy of the system, 
\bq
\lb{1.10}
\frac{\partial^2{\cal{L}}(x, \dot{x}, \ddot{x})}{\partial \ddot{x}^2} \not= 0,
\eq
so the inverse $\ddot{x} = A(X_1, X_2, P_2)$ of Eq.(\ref{1.7}) exists}. The above considerations can be easily generalized to systems with even higher order 
time derivatives  \cite{Woodard15}.

Clearly, with the above theorem one can see that any higher derivative theory of gravity with the Lorentz invariance (LI) and the 
non-degeneracy condition is not stable.  
Taking the above point of view 
 into account,  recently extensions of scalar-tensor theories were investigated by evading the  Ostrogradsky instability \cite{LN16,RM16,Achour}.

Another way to evade  Ostrogradsky's theorem is to break LI in the UV and include only high-order spatial derivative terms in the Lagrangian, while still keep
the time derivative terms to the second order. This is exactly what  Ho\v{r}ava did recently  \cite{Horava}. 

It must be emphasized  that this has to be done with great care. First, LI is one of the fundamental principles of modern physics and   strongly supported by observations.
In fact, all the experiments carried out so far are consistent with it  \cite{Liberati13}, and no evidence to show that such a symmetry
must be  broken at certain energy scales, although  the constraints in  the gravitational sector are much weaker than those in 
the matter sector    \cite{LZbreaking}. Second, the breaking of LI can have significant effects on the low-energy physics  through
the interactions between gravity and matter, no matter how high the scale of symmetry breaking is \cite{Collin04}. Recently, 
it was proposed  a mechanism of SUSY breaking by coupling a Lorentz-invariant supersymmetric matter sector to 
non-supersymmetric gravitational interactions with Lifshitz scaling, and shown that it can lead to a consistent   Ho\v{r}ava gravity \cite{PT14} \footnote{
A supersymmetric version of  Ho\v{r}ava gravity has not been successfully constructed, yet \cite{Xue,Redigolo,PS}.}.
Another scenario   is to go beyond the perturbative realm, so that strong interactions will take over at an intermediate scale (which is in between
the Lorentz violation   and  the infrared   (IR) scales) and accelerate   the renormalization group (RG) flow quickly to the LI fixed point  in the IR  \cite{BPSd,KS,Afshordi}.

With the above in mind,   in this article we shall give an updated review of   Ho\v{r}ava gravity. Our emphases will be on: (i) {\em the self-consistency of the theory},
such as free of ghosts and instability; (ii) {\em consistency  with experiments}, mainly the solar system tests  and cosmological observations; and
(iii) {\em predictions}. One must confess  that this is not an easy task, considering the fact that
the field has been extensively developed in the past few years and there have been various extensions of  Ho\v{r}ava's original {\em minimal  theory}
 (to be defined soon below)\footnote{Up to the moment 
of writing this review,  Ho\v{r}ava's seminal paper \cite{Horava}  has been already cited about 1400 times, see, for example, 
https://inspirehep.net/search?p=find+eprint+0901.3775.}. So, one way or another
one has to make a choice on which subjects that should be included in a brief review,  like the current one. 
Such a choice clearly contains the reviewer's bias. In addition, in this review we do not intend to exhaust all  the relevant articles even within 
the chosen subjects, as in the information era, one can simply find them, for example, from the list of the citations of   Ho\v{r}ava's paper 
\footnote{A more complete list of articles concerning  Ho\v{r}ava gravity can be found from the citation list of Ho\v{r}ava's seminal paper \cite{Horava}:
  https://inspirehep.net/search?p=find+eprint+0901.3775.}. With all these reasons, I would like first to offer my sincere thanks and apologies  to whom his/her work is not mentioned in this review. 
In addition,  there have already existed a couple of excellent  reviews on   Ho\v{r}ava gravity and its applications to cosmology and astrophysics
\cite{Muk,BC,TPS,WSV,Visser11,Padilla,Clifton}, including the one by   Ho\v{r}ava himself \cite{Hreview}. Therefore, for the  works prior to these reviews, the readers
are  strongly  recommended to them for details. 

The rest part of the review is organized as follows: In the next section (Sec. II), we first give a brief introduction to  the gauge symmetry that Ho\v{r}ava gravity adopts and 
the general form of the action that can be constructed under such a symmetry. Then, we state clearly the problems with this incarnation. To solve these problems, various
modifications have been proposed. In this review,  we introduce four of them, respectively,  in Sec. II.A - D,  which have been most intensively studied so far. At the end of this section (Sec. II.E), 
we consider the covariantization of these models, which can be considered as the IR limits of the corresponding versions of Ho\v{r}ava gravity. In Sec. III, we present the recent developments
of universal horizons and black holes, and discuss the corresponding thermodynamics, while in Sec. IV, we discuss the 
non-relativistic gauge/gravity duality, by paying particular attention on spacetimes with  Lifshitz symmetry.  In Sec. V, we consider the quantization of 
Ho\v{r}ava gravity, and summarize the main results obtained so far in the literature. These studies can be easily generalized to other gravitational theories with broken LI. 
The review is ended in Sec. VI, in which we list some open questions of Ho\v{r}ava's quantum  gravity and 
present some concluding remarks. An appendix is also included, in which we give a brief introduction to Lifshitz scalar theory.

Before proceeding to the next section, let us mention some (well studied) theories of QG. These include
string/M-Theory  \cite{stringa,stringb,stringc}, Loop Quantum Gravity (LQG) \cite{LQGa,LQGb,LQGc,LQGd} \footnote{It is interesting to note that big bang
singularities have been intensively studied in Loop Quantum Cosmology (LQC) \cite{LQCc}, and a large number of cosmological models 
have been considered \cite{LQCSb}.
In all of these models big bang singularity is resolved by quantum gravitational effects in the deep Planck regime. Similar conclusions are also obtained for
black holes \cite{LQGBHd}.}, 
Causal Dynamical  Triangulation (CDT) \cite{CDTs},  and Asymptotic Safety \cite{AS,ASa}, to name only few of them. For more details, see  \cite{Bojowald15}.
However, our understanding on each of them  is still highly limited. In particular we do not know the relations among them (if there exists any),  and
more importantly, if any of them  is {\em the theory} we have been looking for \footnote{One may never be able to prove truly that a theory is correct, but rather disprove
or more accurately constrain a hypothesis \cite{Popper}. The history of science tells us that this has been the case so far.}.  
One of the main reasons   is the  lack of  experimental evidences.
This is understandable, considering the fact that quantum gravitational  effects are  expected to become important only at the Planck  scale, which currently  is
well above the range of any man-made terrestrial experiments. However,   the situation has   been changing recently  with the arrival of precision cosmology 
\cite{KW14a,KW14b,KW14c,KW14d,KW14e,KW14f,stringCa,stringCb}. In particular,  the inconsistency of the theoretical predictions   
 with current observations, obtained  by using the deformed algebra  approach in the framework of LQC \cite{deformed2},  has    shown that cosmology 
has indeed already entered an era in which quantum theories  of gravity can be  tested directly by observations. 

\section{ Ho\v{r}ava Theory of Quantum Gravity}
\renewcommand{\theequation}{2.\arabic{equation}} \setcounter{equation}{0}

According to our current understanding, space and time are quantized in the deep Planck regime, and a continuous spacetime  only emerges later as a classical limit
of QG  from some discrete substratum. Then, since the LI is a continuous symmetry of spacetime, it may not exist quantum mechanically, and instead 
 emerges  at the low energy physics. Along this line of arguing, it is not unreasonable to assume that LI is broken in the UV but recovered later in the IR. Once LI is
 broken, one can include only high-order spatial derivative operators into the Lagrangian, so the UV behavior can be improved, while  the time derivative operators
 are still kept to the second-order, in order to evade Ostrogradsky's ghosts. This was precisely  what   Ho\v{r}ava  did \cite{Horava}. 
 
 Of course, there are many ways to break LI. But,  Ho\v{r}ava chose to break it by considering anisotropic scaling between time and space, 
 \bq
\lb{2.1}
t \rightarrow b^{-z} t,\;\;\; x^i \rightarrow b^{-1}{x'}^i,\; (i = 1, 2, ..., d)
\eq
where $z$ denotes the dynamical critical exponent, and LI requires $z = 1$, while power-counting renomalizibality requires $ z \ge d$, where $d$ denotes the spatial dimension of the
spacetime  \cite{Horava,Visser,VisserA,AH07,FIIKa,FIIKb}. In this review we mainly consider spacetimes  with $d = 3$ and take the minimal value 
$z = d$, except for particular considerations. Whenever this happens, we shall make specific notice.    Eq.(\ref{2.1})  is a reminiscent of Lifshitz's scalar fields in condensed
matter physics  \cite{Lifshitz,Lifshitz2}, hence  in the literature Ho\v{r}ava gravity  is also  called  the
Ho\v{r}ava-Lifshitz  (HL) theory. With the scaling of Eq.(\ref{2.1}), the time and space have,   respectively, the dimensions \footnote{In this review we will  measure
canonical dimensions of all objects in the unities of spatial momenta $k$.},
 \bq
\lb{2.2}
[t] = - z, \quad \left[x^i\right] = -1.
\eq
Clearly, such a scaling breaks
explicitly the LI and hence $4$-dimensional diffeomorphism
invariance. Ho\v{r}ava assumed that it is broken only down to the level
\begin{equation}
\lb{2.3}
 t \to  \xi_0(t), \quad {x}^i \to  \xi^i\left(t, x^k\right),
\end{equation}
so the spatial diffeomorphism still remains. The above symmetry is often referred as to {\em the  foliation-preserving diffeomorphism},  denoted  by
Diff(M, ${\cal{F}}$).  To see how gravitational fields transform under the above diffeomorphism, let us first introduce the  Arnowitt-Deser-Misner  (ADM)
variables  \cite{ADM}, 
\begin{equation}
\lb{2.4}
\left(N, N^i, g_{ij}\right),  
\end{equation}
where $N, \; N^i$ and $g_{ij}$, denote, respectively, the lapse function, shift vector, and 3-dimensional metric of the leaves $t = $ constant \footnote{In the ADM decomposition, 
the line element $ds$ is given by $ds^2 = g_{\mu\nu}dx^{\mu}dx^{\nu}$ with the 4-dimensional metrics $g_{\mu\nu}$ and $g^{\mu\nu}$ being given by \cite{ADM}, 
\bq
\lb{ADM}
g_{\mu\nu} = \left(\begin{matrix}
-N^2 + N_iN^i & N_i\cr
N_i & g_{ij}\cr
\end{matrix}
\right), \;
g^{\mu\nu} = \left(\begin{matrix}
-\frac{1}{N^{2}} & \frac{N^{i}}{N^2}\cr
 \frac{N^{i}}{N^2} & g^{ij} -  \frac{N^{i}N^j}{N^2}\cr
\end{matrix}
\right).
\eq
But, in Ho\v{r}ava gravity the line element $ds$ is not necessarily given by this relation [For example, see Eqs.(\ref{2.25a}) and (\ref{2.25b})].
 Instead, one can simply consider
$\left(N, N^i, g_{ij}\right)$ as the fundamental quantities that describe the quantum gravitational field of Ho\v{r}ava gravity, and their relations to the macroscopic quantities, such as $ds$, 
will emerge in the IR limit.}. Under the rescaling (\ref{2.1})
   $N, \; N^{i}$ and $g_{ij}$ are assumed to scale,
 respectively,  as  \cite{Horava},
 \bq
 \lb{2.5}
  N \rightarrow  N ,\;\;\;  N^{i}
\rightarrow b^{2} N^{i},\;\;\; g_{ij} \rightarrow g_{ij},
 \eq
 so that their dimensions are 
 \begin{equation}
\lb{2.6}
 {N} = 0, \quad \left[N^i\right] = 2, \quad \left[g_{ij}\right] = 0.  
\end{equation}
Under the Diff($M, \; {\cal{F}}$), on the other hand, they transform as,
\bqn
\lb{2.7}
\delta{N} &=& \xi^{k}\nabla_{k}N + \dot{N}\xi_0 + N\dot{\xi_0},\nb\\
\delta{N}_{i} &=& N_{k}\nabla_{i}\xi^{k} + \xi^{k}\nabla_{k}N_{i}  + g_{ik}\dot{\xi}^{k}
+ \dot{N}_{i}\xi_0 + N_{i}\dot{\xi_0}, \nb\\
\delta{g}_{ij} &=& \nabla_{i}\xi_{j} + \nabla_{j}\xi_{i} + \xi_0\dot{g}_{ij},  
\eqn
where  $N_{i} \equiv g_{ij}N^{j}$, and  in writing the above we had assumed that $\xi_0(t)$ and $\xi^k\left(t, x^i\right)$ are small, so that only their linear terms appear. 
Once we know the transforms (\ref{2.7}), we can construct the basic operators of the fundamental variables (\ref{2.4}) and their derivatives,
which turn out to be \footnote{Note that with the  general diffeomorphism, $x^{\mu} \to   \xi^{\mu}\left(t, x^k\right)$, the fundamental quantity 
is the  Riemann tensor $R^{\mu}_{\;\;\nu\alpha\beta}$.}, 
\bq
\lb{2.8}
R_{ij}, \;\;\; K_{ij}, \;\;\;  a_i,  \;\;\; \nabla_i,
\eq
where $a_i \equiv N_{,i}/N$,  and $ \nabla_i$ denotes the covariant derivative with respect to $g_{ij}$, while 
 $R_{ij}$ is the 3-dimensional Ricci tensor constructed from the 3-metric $g_{ij}$ and $g^{ij}$ where $g_{ij}g^{ik} = \delta^{k}_{j}$. 
 $K_{ij}$ denotes the extrinsic curvature tensor of the leaves $t$= constant, defined as
 \bq
\lb{2.9}
K_{ij} \equiv  \frac{1}{2N}\left(-\dot{g}_{ij} + \nabla_{i}N_{j} + \nabla_{j}N_{i}\right),
\eq
with $\dot{g}_{ij} \equiv \partial g_{ij} /\partial t$. It can be easily shown that these basic quantities are vectors/tensors under the coordinate transformations
(\ref{2.3}), and have the dimensions,
\bq
\lb{2.10}
 \left[R_{ij}\right] = 2, \;\;\; \left[K_{ij}\right] = 3, \;\;\; \left[a_i\right] = 1, \;\;\; \left[\nabla_i\right] = 1.
\eq
 
With the basic blocks of Eq.(\ref{2.8}) and their dimensions, we can build scalar operators  order by order, so the total Lagrangian 
will finally take the form,
\bq
\lb{2.11}
{\cal{L}}_g = \sum_{n=0}^{2z}{{\cal{L}}^{(n)}_g\left(N, N^i, g_{ij}\right)},
\eq
where ${\cal{L}}_g^{(n)}$ denotes the part of the Lagrangian that contains operators of the nth-order only. In particular,  to each order of $[k]$, we have the 
following independent terms that are all scalars under the transformations of the
foliation-preserving diffeomorphisms (\ref{2.3}) \cite{ZSWW},
\bqn
\lb{2.12}
&& [k]^{6}: K_{ij}K^{ij},\; K^{2},\; R^{3}, \;   RR_{ij}R^{ij} , \; R^{i}_{j}R^{j}_{k}R^{k}_{i},\;  \left(\nabla{R}\right)^{2}, \nb\\
&& ~~~~~~~~ \left(\nabla_{i}R_{jk}\right) \left(\nabla^{i}R^{jk}\right),\;  \left(a_{i}a^{i}\right)^{2}R , \;
  \left(a_{i}a^{i}\right)\left(a_{i}a_{j}R^{ij}\right),\nb\\
&& ~~~~~~~~ \left(a_{i} a^{i}\right)^{3}, a^{i}\Delta^{2}a_{i},  \left(a^{i}_{\;\;i}\right) \Delta{R},  ...,\nb\\
&& [k]^{5}: \;  K_{ij}R^{ij},\; \epsilon^{ijk}R_{il}\nabla_{j}R^{l}_{k},\;  \epsilon^{ijk}a_{i}a_{l}\nabla_{j}R^{l}_{k}, \nb\\
&& ~~~~~~~~  a_{i}a_{j} K^{ij}, \; K^{ij}a_{ij},\;    \left(a^{i}_{\;\;i}\right)K, \nb\\
&& [k]^{4}: \; R^{2},\; R_{ij}R^{ij},\; \left(a_{i}a^{i}\right)^{2},  \; \left(a^{i}_{\;\;i}\right)^{2},\;  \left(a_{i}a^{i}\right)a^{j}_{\;\;j},\nb\\
&&  ~~~~~~~~  a^{ij}a_{ij},\;  \left(a_{i}a^{i}\right)R,\;  a_{i}a_{j}R^{ij},\;  Ra^{i}_{\;\;i},   \nb\\
&& [k]^{3}: \; \omega_{3}(\Gamma),\nb\\
&& [k]^{2}: \; R,\;  a_{i}a^{i},  \nb\\
&& [k]^{1}: \; {\mbox{None}},\nb\\
&& [k]^{0}: \; \gamma_{0},
\eqn
where  $\omega_{3}(\Gamma)$ denotes the gravitational Chern-Simons term, $\gamma_{0}$ is a dimensionless constant,
 $\Delta \equiv g^{ij}\nabla_{i}\nabla_{j}$, and
\bqn
\lb{2.13}
 a_{i_{1}i_{2}...i_{n}}  &\equiv& \nabla_{i_{1}} \nabla_{i_{2}}... \nabla_{i_{n}}\ln(N),\nb\\
\omega_{3}(\Gamma) &\equiv& {\mbox{Tr}} \Big(\Gamma \wedge d\Gamma + \frac{2}{3}\Gamma\wedge \Gamma\wedge \Gamma\Big)\nb\\
&=& \frac{e^{ijk}}{\sqrt{g}}\Big(\Gamma^{m}_{il}\partial_{j}\Gamma^{l}_{km} + \frac{2}{3}\Gamma^{n}_{il}\Gamma^{l}_{jm} \Gamma^{m}_{kn}\Big).
\eqn
Here $g={\rm det}\,\left(g_{ij}\right)$ and $\epsilon^{ijk}  \equiv {e^{ijk}}/{\sqrt{g}}$ with $e^{123} = 1$, etc.
Note that in writing Eq.(\ref{2.12}), we had not written down all the sixth  order terms, as they are numerous  and a complete set of it has not been given explicitly \cite{KP13,CES}. 
Then, the general action of the gravitational part (\ref{2.5}) will be the summary of all these terms.  Since time derivative terms only contain in $K_{ij}$, we can see that
 the kinetic part ${\cal{L}}_{K}$ is the linear combination of the  sixth order derivative terms,  
 \bq
\lb{2.14}
{\cal{L}}_{K} = \frac{1}{\zeta^2}\left(K_{ij}K^{ij}- \lambda K^{2}\right),
\eq
where  $\zeta^2$ is the gravitational coupling constant with the dimension 
\bq
\lb{2.14a}
 \left[\zeta^2\right] =  \left[t\right]\cdot \left[x^i\right]^3 + \left[K\right]^2 = -z - 3 + 2z = z-3. 
 \eq
 Therefore, for $z = 3$, 
it is dimensionless,  and the power-counting analysis given between
Eqs.(\ref{1.1}) and (\ref{1.2}) shows that the theory now becomes power-counting renormalizable. 
The parameter  $\lambda$ is another  dimensionless coupling constant, and   LI guarantees it to be one even after radiative corrections are taken into account. But, in Ho\v{r}ava gravity
it becomes a running constant due to the breaking of LI. 

The rest of the Lagrangian  (also called the potential) will be the linear combination of all
the rest terms of Eq.(\ref{2.12}), from which we can see that,
without protection of further symmetries, the total Lagrangian of the gravitational sector is about   100 terms,  which is normally considered very large and
could potentially diminish  the prediction power of the theory. Note that the odd terms given in
Eq.(\ref{2.12}) violate the parity. So, to eliminate them, we can simply require  that parity be conserved.   However, since there are only six such terms, this will not 
reduce the total number of coupling constants significantly.

To further reduce the number of independent coupling constants,  Ho\v{r}ava introduced two additional  conditions,  {\em the projectability and  detailed balance}
\cite{Horava}. The former requires that the lapse function $N$  
be a function of $t$ only,
\bq
\lb{2.15}
N = N(t),
\eq
so that all the terms proportional to $a_i$ and its derivatives will be dropped out. This will reduce considerably  the total number of the independent  terms in
Eq.(\ref{2.11}), considering the fact that $a_i$ has dimension of one only. Thus, to build an operator out of $a_i$  to  the sixth-order,  there will be many independent combinations
of $a_i$ and its derivatives.
However, once the condition (\ref{2.15}) is imposed, all such terms vanish identically, and the total number of the sixth-order terms immediately reduces 
to seven, given exactly by the first seven terms in Eq.(\ref{2.12}).  So, the totally number 
of the independently coupling constants of the theory now reduce to ${\cal{N}}  = 14$, even with the three parity-violated terms, 
\bq
\lb{2.15aa}
K_{ij}R^{ij},\;\;\; \epsilon^{ijk}R_{il}\nabla_{j}R^{l}_{k},\;\;\;
\omega_{3}(\Gamma).
\eq
 It is this version of the HL theory that Ho\v{r}ava referred to as the {\em minimal theory} \cite{Hreview}.  
Note that the projectable condition (\ref{2.15})  is mathematically elegant and appealing. It is preserved by the Diff($M, \; {\cal{F}}$) (\ref{2.3}), and forms an
independent branch of differential geometry \cite{MM03}. 

Inspired by condensed matter systems \cite{Cardy}, in addition to the projectable condition, 
Ho\v{r}ava also assumed that   
the potential part, ${\cal{L}}_{V}$,  can be obtained from a superpotential $W_{g}$ via
the relations \cite{Horava},
\bqn
\lb{2.16}
{\cal{L}}_{V} \equiv w^{2} E_{ij}{\cal{G}}^{ijkl}E_{kl},\;\;\;
E^{ij} \equiv \frac{1}{\sqrt{g}}\frac{\delta{W}_{g}}{\delta{g}_{ij}},
\eqn
where $w$ is a coupling constant, and ${\cal{G}}^{ijkl}$ denotes the generalized DeWitt metric, defined as
\bq
\lb{2.17}
{\cal{G}}^{ijkl} = \frac{1}{2}\big(g^{ik}g^{jl} + g^{il}g^{jk}\big) - \lambda g^{ij}g^{kl},
\eq
where  $\lambda$ is the same coupling constant, as introduced in Eq.(\ref{2.14}), and 
the superpotential $W_{g}$ is given by
\bq
\lb{2.18}
W_{g}  = \int_{\Sigma}{\omega_{3}(\Gamma)} +  \frac{1}{\kappa^{2}_{W}}\int_{\Sigma}{d^{3}x\sqrt{g}(R - 2\Lambda)},
\eq
where $\Sigma$ denotes the leaves of $t=$ constant, $\Lambda$ the cosmological constant, 
and $ \kappa_{W}$  is another coupling constant of the theory. Then, the total Lagrangian 
${\cal{L}}_{g} = {\cal{L}}_{K} -  {\cal{L}}_{V}$,
 contains only five coupling constants, $\zeta, \lambda, w, \kappa_w$ and $\Lambda$. 

Note that the above detailed balance condition  has  a couple of remarkable features \cite{Hreview}:  First, it is in the same spirit of the
AdS/CFT correspondence  \cite{AdSCFTa,AdSCFTb,AdSCFTc,AdSCFTd,AdSCFTe}, where the superpotential is defined on the 3-dimensional leaves,
$\Sigma$, while the gravity is (3+1)-dimensional. Second, in  the non-equilibrium  thermodynamics,  the counterpart of the superpotential $W_{g}$
plays the role of entropy, while the term $E^{ij}$ the entropic forces  \cite{OM,MO}. This might shed   light on the nature of the gravitational forces
 \cite{Verlinde}.

 However, despite of these desired features, this condition leads to several  problems, including  that
the Newtonian  limit does not exist \cite{LMP},  and  the six-order derivative operators are eliminated, so the theory is still not power-counting  renormalizable \cite{Horava}. 
In addition, it is not clear if this symmetry is still respected  by radiative corrections. 

Even more fundamentally,  the foliation-preserving diffeomorphism (\ref{2.3}) allows one more degree of freedom in the gravitational sector, in comparing with that of 
general  diffeomorphism.    As a result,   a spin-0 mode of  gravitons appears. This mode is potentially dangerous and
may cause ghosts and  instability problems, which  lead the constraint algebra dynamically inconsistent   \cite{CNPS,LP,BPS,HKLG}. 

To solve  these problems, various modifications  have been proposed \cite{Muk,BC,TPS,WSV,Visser11,Padilla,Clifton,Hreview}.
In the following we shall briefly introduce only 
four of them, as they have been most extensively studied in the literature so far. These are the ones: (i) with the projectability condition - {\em the minimal theory}
\cite{Horava,SVWa,SVWb}; (ii) with the projectability condition and an extra U(1) symmetry \cite{HMT,daSilva}; (iii) without the projectability condition
but including all the possible terms 
- {\em the healthy extension} \cite{BPSa,BPSb}; and (iv)   with an  extra U(1) symmetry  but without the projectability condition \cite{ZWWS,ZSWW,LMWZ}. 

Before considering each of these models in detail,  some comments on  singularities in Ho\v{r}ava gravity are in order, as they will
appear in all   of these models and shall be faced when we consider  applications of Ho\v{r}ava gravity to    black hole
physics and cosmology \cite{CW10}. 
First,   the nature of singularities of  a given spacetime  in Ho\v{r}ava theory could be quite different from that  in GR, which has the   general diffeomorphism, 
\bq
\lb{2.18a}
\delta x^{\mu} =   \xi^{\mu}\left(t, x^k\right), \; (\mu = 0, 1, 2, 3).
\eq
In GR,  singularities  are divided  into two different kinds:
{\em spacetime  and coordinate
singularities} \cite{ES77}. The former is real and cannot be removed by any  coordinate  transformations
of the type given by Eq.(\ref{2.18a}). The latter is coordinate-dependent, and can be removed by proper coordinate  transformations of the kind (\ref{2.18a}). 
Since the laws of  coordinate  transformations in GR and  Ho\v{r}ava theory are different, it is clear that the nature of singularities are also different. In GR it may be a coordinate 
singularity but in  Ho\v{r}ava gravity it becomes a spacetime singularity. 
Second, two different metrics may represent the same spacetime in GR but in general it is no longer  true in Ho\v{r}ava theory. A concrete example is the Schwarzschild solution
given in the Schwarzschild coordinates,
\bq
\lb{2.18b}
ds^2 = - \left(1 - \frac{2m}{r}\right)dt^2 +    \left(1 - \frac{2m}{r}\right)^{-1}dr^2 + r^2d\Omega^2.
\eq
Making the coordinate transformation, 
\bq
\lb{2.18c}
dt_{PG} = dt + \frac{\sqrt{2mr}}{r - 2m} dr,
\eq
the above metric takes the  Painleve-Gullstrand (PG) form \cite{PG},
\bq
\lb{2.18d}
ds^2 = - dt_{PG}^2 + \left(dr + \sqrt{\frac{2m}{r}}\; dt_{PG}\right)^2  + r^2d\Omega^2.
\eq
In GR we consider  metrics (\ref{2.18b}) and (\ref{2.18d}) as describing  the same spacetime (at least in the region $ r > 2m$), as they are connected by the coordinate transformation
(\ref{2.18c}), which is allowed by the symmetry (\ref{2.18a}) of GR. But this is no longer the case when we consider them  in  Ho\v{r}ava gravity. 
The coordinate transformation (\ref{2.18c}) is not allowed by the foliation-preserving diffeomorphism (\ref{2.3}), and as a result,  they describe two different 
spacetimes  in Ho\v{r}ava theory. In particular, metric (\ref{2.18d}) satisfies the projectability condition, while metric (\ref{2.18b}) does not. So, in Ho\v{r}ava theory they belong to the 
two  completely different  branches, with or without the projectability condition.  Moreover, in GR the metric
\bq
\lb{2.18e}
ds^2 = - \left(1 - \frac{2m}{r}\right)dv^2 +   2dvdr     + r^2d\Omega^2,
\eq
describes the same spacetime as  those of metrics (\ref{2.18b}) and (\ref{2.18d}), but it does not belong to any of the two branches of Ho\v{r}ava theory, 
because $v=$ constant  hypersurfaces do not define a ($3+1$)-dimensional foliation, a fundamental  requirement of Ho\v{r}ava gravity.   Third, because of the difference between the two kinds of coordinate
transformations, the global structure of a given spacetime is also different in GR and  Ho\v{r}ava gravity \cite{GLLSW}. For example, the maximal extension of the Schwarzschild solution was 
achieved when it is written in the Kruskal coordinates \cite{HE73},
\bq
\lb{2.18f}
ds^2 = - \frac{e^{-r/2m}}{r}dUdV     + r^2(U,V) d\Omega^2.
\eq
But the coordinate transformations that bring metric (\ref{2.18b}) or (\ref{2.18d}) into this form are not allowed by the  foliation-preserving diffeomorphism (\ref{2.3}). 
For more details, we refer readers to \cite{CW10,GLLSW}.

\subsection{The Minimal Theory}

If we only impose the parity  and projectability  condition (\ref{2.15}), the total action for the gravitational sector can be cast in the form  \cite{SVWa,SVWb},
\bqn 
\lb{action}
S_g = \zeta^2\int dt d^{3}x N \sqrt{g} \left({\cal{L}}_{K} -
{\cal{L}}_{{V}}\right),
 \eqn
where  ${\cal{L}}_{K}$  is given by Eq.(\ref{2.14}),
while the potential part takes the form,
 \bqn \lb{2.19}
{\cal{L}}_{{V}} &=& 2\Lambda - R + \frac{1}{\zeta^{2}}
\left(g_{2}R^{2} +  g_{3}  R_{ij}R^{ij}\right)\nb\\
& & + \frac{1}{\zeta^{4}} \left(g_{4}R^{3} +  g_{5}  R\;
R_{ij}R^{ij}
+   g_{6}  R^{i}_{j} R^{j}_{k} R^{k}_{i} \right)\nb\\
& & + \frac{1}{\zeta^{4}} \left[g_{7}R\nabla^{2}R +  g_{8}
\left(\nabla_{i}R_{jk}\right)
\left(\nabla^{i}R^{jk}\right)\right].
 \eqn
Here $\zeta^{2} = 1/{16\pi G}$,  and  $g_{n}\,
(n=2,\dots 8)$  are all dimensionless  coupling constants. Note that,  without loss of the generality,  in writing Eq.(\ref{2.19}) the coefficient in the front of $R$ was set to 
$-1$, which   can be realized  by rescaling the time and space
coordinates \cite{SVWb}. As mentioned above, Ho\v{r}ava referred to this model as  the {\em minimal theory} \cite{Hreview}. 

In the IR, all the high-order curvature terms (with coefficients $g_n$'s) drop out, and the total action 
reduces  to the Einstein-Hilbert action, provided that the coupling constant $\lambda$ flows to its relativistic limit $\lambda_{GR}=1$ in the IR. 
This has not been shown  in the general case. But, with only the three coupling constants ($\zeta, \; \Lambda, \lambda$), it was found that the Einstein-Hilbert action 
with $\Lambda = 0$ is  an attractor in the phase space of  RG flow \cite{CRS}. In addition, RG trajectories with a tiny positive cosmological constant 
also come with a value of $\lambda$  that is compatible with experimental constraints.
   
 To study the stability of the theory, let us consider the linear perturbations of the Minkowski background (with $\Lambda = 0$),
 \bqn
 \lb{2.19a}
 N =1, \quad N_i = \partial_i B, \quad g_{ij} = e^{-2\psi}\delta_{ij}.
 \eqn
 After integrating out the $B$ field, the action upto the quadratic terms of $\psi$ takes the form  \cite{WM10}, 
 \bq
 \lb{2.19b}
 s_{g}^{(2)} = \zeta^2 \int{dt dx^3 \left(- \frac{\dot{\psi}^2}{c_{\psi}^2} - \psi\left(1 + \alpha_1\partial^2 +  \alpha_2\partial^4\right)\partial^2\psi\right)},
 \eq
where $\partial^2 \equiv \delta^{ij}\partial_i\partial_j,\; c_{\psi}^2 \equiv -(\lambda-1)/(3\,\lambda-1)$, $\alpha_1 \equiv (8g_2 + 3g_3)/\zeta^2$ and $\alpha_2 \equiv - (8g_7 - 3g_8)/\zeta^2$. Clearly,
to avoid ghosts we must assume that $c_{\psi}^2 < 0$, that is,
\bq
 \lb{2.19c}
(i)\;  \lambda > 1  \quad {\mbox{or}} \quad (ii)\; \lambda < \frac{1}{3}. 
 \eq
However, in these intervals the scalar mode is not stable in the IR \cite{SVWb,WM10} \footnote{ Stability of the scalar mode with the projectability condition was also
considered in \cite{BS}. But, it was found that it exists for all the value of $\lambda$. This is because   the detailed balance condition was also imposed in \cite{BS}.}. 
This can be seen easily from  the equation of motion of $\psi$ in the momentum space,  
\bq
 \lb{2.19d}
\ddot{\psi}_k + \omega_k^2  \psi_k = 0,   
 \eq
where $\omega_k^2  \equiv c_{\psi}^2\left(1 - \alpha_1 k^2 +  \alpha_2 k^4\right)k^2$. In the  intervals of Eq.(\ref{2.19c}), we have   
$c_{\psi}^2 < 0$ in the IR, so $\omega_k^2$ also becomes negative, that is, the theory suffers tachyonic instability. In the UV and intermediate regimes, it can be stable by 
properly choosing the coupling constants of the high-order operators
$g_n (n = 2, 3, 7, 8)$.  Note that each of these terms is subjected to radiative corrections. It would be very interesting to show that the scalar mode  is
still stable, even after such corrections are taken into account.  

 On the other hand, from  Eq.(\ref{2.19b}) it can be seen that there are two particular values of $\lambda$ that make the above analysis invalid, one is $\lambda =1$ and the other is 
$\lambda = 1/3$.  A more careful analysis of these two cases  shows that  the equations for $\psi$ and $B$ degenerate into elliptic differential equations, so  the scalar mode is 
no longer dynamical.  As a result, the Minkowski spacetime in these two cases are stable.

It is also interesting to note that the de  Sitter spacetime in this minimal theory  is stable \cite{WW10,HWW}. See also a recent study of the issue in a closed FLRW universe \cite{MFM}.

 In addition,  this minimal   theory still suffers the strong coupling problem \cite{KA,WW10}, so the power-counting analysis presented above becomes invalid. It must be noted that this 
does not necessarily imply the loss of predictability: if the theory is renormalizable, all coefficients of
infinite number of nonlinear terms can be written in terms of finite parameters in the action, as several well-known
theories with strong coupling (e.g., \cite{Pol}) indicate. However, because of the breakdown of the (naive) perturbative
expansion, we need to employ nonperturbative methods to analyze the fate of the scalar graviton in the
limit. Such an analysis was performed in \cite{Muk} for spherically symmetric, static, vacuum configurations and 
was shown that the limit is continuously connected to GR.   A similar consideration for cosmology was given in \cite{Izumi:2011eh,GMW}, where a fully
nonlinear analysis of superhorizon cosmological perturbations was carried out, and  was  shown that the limit $\lambda \to 1$  is continuous
and that GR is recovered.   This may be considered as an analogue of the Vainshtein effect  first found in massive gravity \cite{Vainshtein:1972sx}. 

With the projectability condition, the Hamiltonian constraint becomes global, from which it was shown that  a component which behaves like dark matter 
emerges as an ``integration constant'' of dynamical equations and momentum constraint equations \cite{Muk09}.   

Cosmological perturbations  in this version of theory has been extensively studied \cite{CHZ,WM10,WWM,Wang10,KUY,GKS,CB11,MFM}, and are found consistent 
with current observations. 

In addition, spherically symmetric spacetimes without/with the presence of matter 
 were also investigated  \cite{IM09,GPW,Satheesh}, and was found that the solar system 
tests can be satisfied by properly choosing the coupling constants of the theory.

 \subsection{With   Projectability $\&$ U(1) Symmetry }  
 
 As mentioned above,  the problems plagued in Ho\v{r}ava gravity are closely related to the existence of the spin-0 graviton. Therefore, if it is
 eliminated, all the problems should be cured. This can be done, for example, by imposing  extra symmetries, which was precisely what Ho\v{r}ava and
Melby-Thompson (HMT) did in \cite{HMT}. HMT introduced   an extra local U(1) symmetry,  so that    
the total symmetry of the theory now is enlarged  to,
\bq
\lb{2.20}
 U(1) \ltimes {\mbox{Diff}}(M, \; {\cal{F}}).
\eq
The extra U(1) symmetry is realized by introducing two auxiliary fields,
the $U(1)$ gauge field $A$ and the Newtonian pre-potential $\varphi$ \footnote{In the original paper of HMT, the  Newtonian pre-potential   was denoted by
$\nu$ \cite{HMT}. In this review, we shall replace it by $\varphi$, and reserve $\nu$ for other use.}.
Under this extended symmetry,   the special status of time  maintains,  so that the anisotropic scaling (\ref{2.1})
  is still  valid, whereby the UV behavior of the theory can be considerably improved.  On the other hand, 
under the local $U(1)$ symmetry,  the  fields 
 transform as
\bqn
\lb{2.21}
\delta_{\alpha}A &=&\dot{\alpha} - N^{i}\nabla_{i}\alpha,\;\;\;
\delta_{\alpha}\varphi = - \alpha,\nb\\ 
\delta_{\alpha}N_{i} &=& N\nabla_{i}\alpha,\;\;\;
\delta_{\alpha}g_{ij} = 0 = \delta_{\alpha}{N},
\eqn
where $\alpha\left(t, x^k\right)$ is   the generator  of the local $U(1)$ gauge symmetry. 
Under the Diff($M,  {\cal{F}}$), the auxiliary fields $A$ and $\varphi$ transform as,
\bqn
\lb{2.22}
\delta{A} &=& \zeta^{i}\nabla_{i}A + \dot{f}A  + f\dot{A},\nb\\
\delta \varphi &=&  f \dot{\varphi} + \zeta^{i}\nabla_{i}\varphi,
\eqn
while $N, N^i$ and $g_{ij}$  still transform as those given by Eq.(\ref{2.7}). 
With this enlarged symmetry, the spin-0 graviton is indeed eliminated
\cite{HMT,WW11}.

At the initial, it was believed  that the $U(1)$ symmetry can be realized only when  the coupling constant $\lambda$  takes its relativistic value   $\lambda = 1$.
This was very encouraging, because  it is the deviation of $\lambda$ from one that causes all the   problems, including  ghost, instability and  strong coupling
\footnote{Recall that  in the  relativistic case,  $\lambda = 1$ is protected by the diffeomorphism,  
$\tilde{x}^{\mu} =   \tilde{x}^{\mu}(t, x^k),\; (\mu = 0, 1, 2, 3)$.  With this symmetry,  $\lambda$ remains this value even after the radiative corrections
are taken into account.}. However, this claim  was soon challenged, and shown  that the introduction of the   Newtonian pre-potential is so strong that action with
$\lambda \not=1$ also has the local $U(1)$ symmetry  \cite{daSilva}. It is remarkable that the spin-0 graviton is still eliminated even with
an arbitrary value of $\lambda$  first by considering linear perturbations  in Minkowski and de Sitter spacetimes \cite{daSilva,HW11}, and then
 by analyzing the Hamiltonian structure of the theory  \cite{Kluson11}. 

The general action for the gravitational sector now takes the form \cite{HW11},
 \bqn \lb{2.23}
S_g &=& \zeta^2\int dt d^{3}x N \sqrt{g} \Big({\cal{L}}_{K} -
{\cal{L}}_{{V}} +  {\cal{L}}_{{\varphi}} +  {\cal{L}}_{{A}} +  {\cal{L}}_{{\lambda}}\Big),\nb\\
 \eqn
where ${\cal{L}}_{K}$ and  ${\cal{L}}_{{V}}$ are given by Eqs.(\ref{2.14}) and (\ref{2.19}), and 
 \bqn \lb{2.24}
{\cal{L}}_{\varphi} &\equiv&\varphi {\cal{G}}^{ij} \Big(2K_{ij} + \nabla_{i}\nabla_{j}\varphi\Big),\nb\\
{\cal{L}}_{A} &\equiv&\frac{A}{N}\Big(2\Lambda_{g} - R\Big),\nb\\
{\cal{L}}_{\lambda} &\equiv& \big(1-\lambda\big)\Big[\big(\nabla^{2}\varphi\big)^{2} + 2 K \nabla^{2}\varphi\Big],\nb\\
{\cal{G}}_{ij} &\equiv& R_{ij} - \frac{1}{2}g_{ij}R + \Lambda_{g} g_{ij}.
 \eqn
Here  $\Lambda_{g}$ is another    coupling constant, and has the same dimension of $R$. Note that the potential ${\cal{L}}_{{V}}$ takes the same form
as that given  in the case without the extra U(1) symmetry. This is because  $g_{ij}$ does not change under the local U(1) symmetry, as it can be seen from
Eq.(\ref{2.21}). So, the most general form of ${\cal{L}}_{{V}}$ is  still given by Eq.(\ref{2.19}).

Note that the strong coupling problem no longer  exists in the gravitational sector, as  the spin-0 graviton now is eliminated. 
However, when coupled with matter,  it will appear again  for processes with energy higher than \cite{HW11,LWWZ}, 
\bq
\lb{2.25}
\Lambda_{\omega} \equiv M_{pl} \left(\frac{M_{pl}}{{\cal{C}}}\right)^{3/2} \; |\lambda - 1|^{5/4},
\eq
where  $M_{pl}$ denotes the Planck mass, and generically ${\cal{C}} \ll M_{pl}$. To solve this problem, one way is to introduce a new energy scale 
$M_{*}$  so that $M_{*} < \Lambda_{\omega}$,  as Blas, Pujolas and Sibiryakov first introduced in the nonprojectable case \cite{BPSc}. This is reminiscent of the case in string theory
where the string scale is introduced just below the Planck scale, in order to avoid strong coupling \cite{stringa,stringb,stringc}. In the rest of this review, it 
will be   referred to as the BPS mechanism.   The main ideas are
the following:  before the strong coupling energy $\Lambda_{\omega}$  is reached [cf. Fig. 1], the sixth order derivative operators  become dominant, so the scaling
law of a physical quantity for process with $E > M_{*}$  will follow Eq.(\ref{2.1}) instead of the  relativistic one ($z =1$).  Then, with such anisotropic scalings, it can be shown
that  all the  nonrenornalizable terms (with $z = 1$) now become either strictly  renormalizable or  supperrenormalizable \cite{Pol}, whereby  the  strong  coupling problem is resolved. 
For more details, we refer readers to \cite{LWWZ,BPSc}.

 \begin{figure}[tbp]
\centering
\includegraphics[width=8cm]{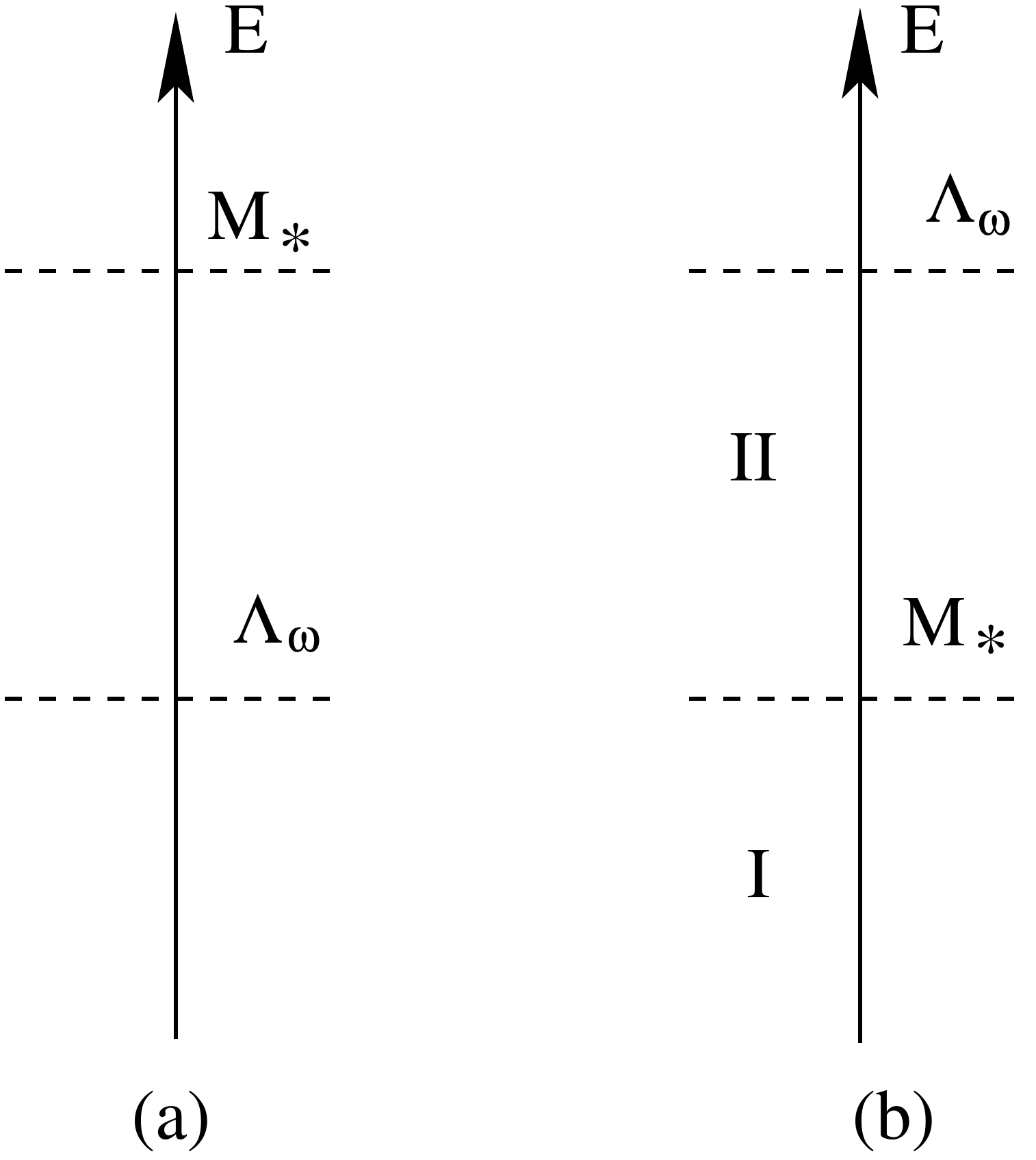}
\caption{The energy scales: (a) $\; \Lambda_{\omega} <  M_{*}$ and (b) $\; \Lambda_{\omega} > M_{*}$.
In Case (a), the theory becomes strong coupling once the energy of a system reaches the strong coupling energy $\Lambda_{\omega}$.
In Case (b), the high-order  operators suppressed by $M_*$  become important before    the system reaches the strong coupling energy
$\Lambda_{\omega}$, whereby the (relativistic) IR scaling of the system is taken over by the anisotropic scaling (\ref{2.1}).}
\label{fig1}
\end{figure}

It should be noted that, in order for the  mechanism to work, the price to pay is that now $\lambda$ cannot be exactly equal to one, as one can see from Eq.(\ref{2.25}). 
In other words,  the theory cannot reduce precisely  to GR  in the IR. However, since GR has achieved great success in low energies, $\lambda$ cannot be significantly 
different from one in the IR, in order for the theory to be consistent with  observations.

In addition, the BPS mechanism cannot be applied to {\em the  minimal theory} presented in the last subsection,  because  the condition $M_{*} < \Lambda_{\omega}$, 
together with the one that instability cannot occur within the age of the universe, 
requires fine-tuning $|\lambda - 1| < 10^{-24}$,  as shown explicitly in \cite{WW10}. However, in the current setup (with any $\lambda$), the Minkowski spacetime is stable, 
so such a fine-tuning does not exist. 

Static and spherically symmetric spacetimes were considered in \cite{HMT,LMW,AP10,GSW,GLLSW,BLW11,RLPV,LMWZ}, and solar system tests were in turn investigated. 
In particular,  HMT found that GR can be recovered  in the IR if the lapse function ${\cal{N}}$ of GR is related to the one $N$ of Ho\v{r}ava gravity  and the gauge field $A$ via
the relation, ${\cal{N}} = N - A$ \cite{HMT}. This can be further justified by the considerations of geometric interpretations of the gauge field $A$, in which the local U(1) symmetry
was found in the first place.  By requiring that the line element is invariant not only  under the Diff(M, ${\cal{F}}$) but also under the local U(1) symmetry, the authors in \cite{LMW}
found that the lapse function ${\cal{N}}$ and the shift vector  ${\cal{N}}^i$ of GR should be given by ${\cal{N}} = N - \upsilon(A - {\cal{A}})/c^2$ and 
 ${\cal{N}}^i =  N^i+N\nabla^{i}\varphi$, where $\upsilon$ is a 
 dimensionless coupling constant,  and 
 \bq
 \lb{2.25aa}
 {\cal{A}} \equiv - \dot{\varphi} + N^i\nabla_{i}\varphi + \frac{1}{2} N(\nabla_{i}\varphi)^2.
 \eq
Then,  it was found that the theory is  consistent with the solar system  tests for both $\lambda=1$ and $\lambda\not=1$, provided that
 $|\upsilon-1|<10^{-5}$. To couple with matter fields, in \cite{LMWZ} the authors considered a universal coupling between matter and the HMT theory via
 the effective metric $\gamma_{\mu\nu}$,
 \bqn
 \lb{2.25a}
 ds^2 &\equiv &\gamma_{\mu\nu}dx^{\mu}dx^{\nu}  \nb\\
 &=&   -c^2{\cal{N}}^2 dt^2 + \gamma_{ij}\left(dx^i + {\cal{N}}^idt\right)\left(dx^j + {\cal{N}}^j dt\right), \nb\\
 \eqn
 where
 \bqn
\lb{2.25b}
{\cal{N}} &\equiv& \left(1 - a_1\sigma\right)N, \quad  {\cal{N}}_i = N_i + N \nabla_i\varphi, \nb\\
 \gamma_{ij} &\equiv&  \left(1 - a_2\sigma\right)^2 g_{ij}, \quad \sigma \equiv \frac{A -  {\cal{A}}}{N}.
\eqn
Here $a_1 (\equiv \upsilon)$ and $a_2$ are two coupling constants. It can be shown that  the effective metric (\ref{2.25a}) is invariant under the enlarged symmetry (\ref{2.20}).
The matter is minimally coupled with respect to the effective metric $\gamma_{\mu\nu}$ via the relation,
 \bqn
\lb{2.25c}
S_{M} = \int{dt d^3x {\cal{N}} \sqrt{\gamma} {\cal{L}}_{M}\left({\cal{N}}, {\cal{N}}^i, \gamma_{jk}; \psi_n\right)},
\eqn
where $\psi_n$ denotes collectively  matter fields. 
With such a coupling,  in \cite{LMWZ} the authors  calculated explicitly  all  the parameterized  post-Newtonian (PPN) parameters  in terms of the 
coupling constants of  the theory, and showed that the theory satisfies the constraints   \cite{Will} \footnote{Note that in general covariant theory including GR, 
the  term $\mathfrak{B} \equiv \int\frac{\rho'}{|\mathbf{x}-\mathbf{x}'|}(\mathbf{x}-\mathbf{x}')\cdot \frac{d\mathbf{v}'}{dt} d^3x'$ appearing in the component
$h_{tt}$  can be always eliminated by the coordinate transformation $\delta t = \xi^0(t, x^k)$. However, in Ho\v{r}ava gravity, this symmetry is missing, and 
the  term $\zeta_B \mathfrak{B}$ must be included into $h_{tt}$. So, instead of the ten PPN parameters introduced in \cite{Will}, here we have an additonal 
one $\zeta_B$. For more details, we refer readers to \cite{LMWZ}.},
\bqn
\lb{2.25ca}
&& \gamma = 1+(2.1\pm2.3)\times10^{-5},\nb\\
&& \beta = 1+(-4.1\pm7.8)\times10^{-5},\nb\\
&& \alpha_1< 10^{-4},\;\;
\alpha_2 <4\times10^{-7},\;\; 
 \alpha_3 < 4\times10^{-20},\nb\\
&& \xi < 10^{-3},\;\;
\Gamma < 1.5\times10^{-3},\;\;
\zeta_1<2\times10^{-2},\nb\\
&& \zeta_2 < 4\times10^{-5},\;\;
\zeta_3 < 10^{-8},\;\;
\zeta_4 < 6\times10^{-3},
\eqn
obtained by all the current solar system tests, where
\bq
\lb{2.25cb}
\Gamma \equiv 4\beta-\gamma-3-\frac{10}{3}\xi-\alpha_1+\frac{2}{3}\alpha_2-\frac{2}{3}\zeta_1-\frac{1}{3}\zeta_2.
\eq
In particular, one can obtain the same results as those given in GR \cite{Will},
\bqn
\lb{PPNa}
&& \gamma = \beta = 1,\;\;\; \alpha_1 = \alpha_2 = \alpha_3 = \xi = 0,\nb\\
&& \zeta_1 =  \zeta_2 =  \zeta_3 =  \zeta_4 = \zeta_B = 0,
\eqn
 for  
\bq
\lb{2.25d}
(a_1,  a_2) = (1, 0).
\eq
 It is interesting to note that this is exactly the case first considered in HMT \cite{HMT}. 
 
 A  remarkable feature is that the solar system tests impose no constraint
 on the parameter $\lambda$. As a result,  when combined with the  condition for the avoidance of the strong coupling problem, these conditions  do not lead to an upper
 bound on the energy scale $M_{*}$ that suppresses higher dimensional operators in the theory. This is in sharp contrast to other versions of
 Ho\v{r}ava gravity without the U(1) symmetry. It should be noted that  the physical meaning 
of the gauge field $A$ and the Newtonian prepotential $\varphi$ were also studied in \cite{AdsSilva} but with a different coupling with matter fields. 
 
 Inflationary cosmology  was studied in detail in \cite{HWW12}, and found that, among other things,   the FLRW universe is necessarily flat. 
 In the sub-horizon regions, the metric and inflaton are tightly coupled and have the same oscillating frequencies. In the super-horizon regions, the perturbations 
 become adiabatic, and the comoving curvature perturbation is constant.   Both scalar and tensor perturbations are almost scale-invariant, and the spectrum 
 indices are the same as those given in GR, but the ratio of the scalar and tensor power spectra depends on the high-order spatial derivative terms, and can be 
 different from that of GR significantly. Primordial non-Gaussianities of scalar and tensor modes were also studied in \cite{HW12,HWYZ} \footnote{
 It should be noted that such studies were
 carried out when matter is minimally coupled to $g_{\mu\nu}$ \cite{HWW12}, and for the universal coupling (\ref{2.25b}) 
 such studies  have not been  worked out, yet. A preliminary study indicates that a more general coupling might be needed \cite{MWWZ}. }.
 
Note that gravitational collapse of a spherically symmetric object was studied systematically in \cite{GLSW}, by using distribution theory.
 The junction conditions across the surface of a collapsing star were derived under the (minimal) assumption that the junctions must be mathematically meaningful in 
 terms of distribution theory. Lately, gravitational collapse in this setup  was investigated and  various solutions were constructed   \cite{daSilva1,daSilva2,daSilva3}.
 
We also note that  with the U(1) symmetry, the detailed balance  condition can be imposed \cite{BLW}. However, in order to have a healthy IR limit,
it is necessary  to break it softly. This will allow the existence of the Newtonian limit in the IR and meanwhile be power-counting renormalizable in the UV.  Moreover, 
with the detailed balance  condition softly breaking, the number of independent coupling constants can be still  significantly reduced. 
This is particularly the case when we consider  Ho\v{r}ava gravity without the projectability condition but with the U(1) symmetry. Note that, even in the latter, the U(1) symmetry is
crucial in order not to have the problem of power-counting renormalizability in the UV, as shown explicitly in \cite{ZSWW}. In particular,  in the healthy extension to be discussed in the
next subsection  the detailed  balance  condition cannot be imposed even allowing it to be broken softly. Otherwise, it can be shown that the sixth-order operators are eliminated by
this condition, and the resulting theory is not power-counting renormalizable. 

It is also interesting  to note that, using the non-relativistic gravity/gauge correspondence, it was found that this version of Ho\v{r}ava gravity has one-to-one correspondence to 
dynamical Newton-Carton geometry without torsion, and a precise dictionary was built \cite{HO15}.

  \subsection{The Healthy Extension}
  
 Instead of eliminating the spin-0 graviton, BPS chose to live with it and work with the non-projectability condition \cite{BPSa,BPSb},
  \bq
 \lb{2.26}
 N = N\left(t, \; x^k\right). 
 \eq
 Although again we are facing  the problem of a large number of coupling constants, BPS showed that the spin-0 graviton can be stabilized even in the IR.
 This is realized by including the quadratic  term $\left(a_ia^i\right)$ into the Lagrangian,
  \bq
 \lb{2.27}
 {\cal{L}}_V = 2\Lambda - R - \beta_0 a_ia^i +   \sum_{n=3}^{6}{{\cal{L}}_V^{(n)}},
 \eq
 where $\beta_0 $ is a dimensionless coupling constant, and ${\cal{L}}_V^{(n)}$ denotes the Lagrangian built by the nth-order operators only.  Then, in  the IR it can be
 shown that the scalar perturbations can be still described by Eq.(\ref{2.19d}), but now with \cite{BPSa,BPSb}
  \bq
 \lb{2.28}
\omega_k^2  = \frac{\lambda -1}{3\lambda -1} \left(\frac{2}{\beta_0 } -1 \right) k^2  + {\cal{O}}\left(\frac{k^4}{M_*^2}\right),  
 \eq
 where $M_*$ is the energy scale of the theory. While the ghost-free condition still leads to the condition (\ref{2.19c}), because of the presence of the $\left(a_ia^i\right)$ term,
 the scalar mode becomes stable for
 \bq
 \lb{2.29}
0 < \beta_0  < 2.  
 \eq
It is remarkable to note that the stability requires $\beta_0  \not=0$ strictly \footnote{It is interesting to note that in the case $\lambda  = 1/3$ two extra second-class constraints 
appear \cite{BRS,BR16}. As a result, in this case  the spin-0 graviton is eliminated even when $\beta_0   =0$. However, since $\beta_0 $ in general  is subjected to radiative corrections, 
it is not clear  which symmetry   preserves this particular value. In  \cite{Horava}, Ho\v{r}ava  showed that at this fixed point the theory has
a conformal symmetry, provided that the detailed balance condition is satisfied. But, there are two issues related to the detailed balance condition as mentioned before: 
(i) the resulted theory  is no longer  power-countering, as the sixth-order operators are eliminated by this condition; and (ii) in the IR the Newtonian limit  does not exist, 
so the theory is not consistent with observations. For further discussions of the running of the coupling constants in terms of RG flow, see \cite{CRS}.}.  
This will lead to significantly difference from the case $\beta_0  = 0$. The stability of the spin-2 mode 
can be shown by realizing the fact that only the following high-order terms have contributions in the quadratic level of linear perturbations of the Minkowski spacetime \cite{BPSb,KP13},
\bqn
 \lb{2.30}
{\cal{L}}_V^{(n\ge 3)}  &=& \zeta^{-2}\left(\gamma_1 R^2 + \gamma_2 R_{ij}R^{ij}  +  \gamma_3 R\nabla_i a^i +  \gamma_4 a_i\Delta a^i\right)\nb\\
&+&  \zeta^{-4}\left[\gamma_5  \left(\nabla_iR\right)^2 + \gamma_6 \left(\nabla_iR_{jk}\right)^2 +    \gamma_7 \left(\Delta R\right)\nabla_i a^i\right.\nb\\
&& \left.  ~~~~~~~~~ +  \gamma_8 a^i\Delta^2 a_i\right],
 \eqn
 where   $\gamma_n$'s are all dimensionless coupling constants.

 As first noticed by BPS, the most stringent constraints   come from the  preferred frame effects due to Lorentz violation, which require  \cite{BPSb,Will},
\bq
\lb{2.31}
|\lambda - 1| \lesssim 4\times 10^{-7},\;\;\; M_{*} \lesssim 10^{15} \; {\mbox{GeV}}.
\eq
In addition,  the timing of active galactic nuclei \cite{Timing} and gamma ray bursts \cite{GammaRay} require
\bq
\lb{2.32}
M_{*} \gtrsim 10^{10} \sim  10^{11} \; {\mbox{GeV}}.
\eq
To obtain the constraint (\ref{2.32}), BPS used the results from the Einstein-aether theory, as these two theories  coincide in the IR  \cite{Jacob,Jacob13}. 

Limits from binary pulsars were also studied recently, and the most stringent constraints of the theory were obtained \cite{YBYB,YBBY}. However, when the allowed range of the  
preferred frame effects  given by the solar system tests is saturated, the limit for $\lambda$  is still given by Eq.(\ref{2.31}), so the upper bound $M_*$ remains the same. 

It is also interesting to note that  observations of synchrotron radiation from the Crab Nebula seems to require
that the scale of Lorentz violation in the matter sector must be $M_{pl}$,  not $M_*$ \cite{LMS12}. In addition,  
the consistency of the theory with the current observations of gravitational waves from the events GW150914 and GW151226 
were studied recently, and moderate constraints were obtained \cite{YYP}.

Since the spin-0 graviton generically appears in this version of Ho\v{r}ava gravity \cite{Kluson10,DJ11}, strong coupling problem is inevistable \cite{PS09,KP10}. However, as mentioned in the
last subsection,  this can be solved by introducing an energy scale $M_{*}$ that suppresses the high-order operators \cite{BPSc}. When the energy of the system
is higher than $M_*$, these high-order operators become dominant, and the scaling law will be changed from $z = 1$ to  with $z = d$ given by  Eq.(\ref{2.1}). With 
such anisotropic scalings,   all the  nonrenornalizable terms (in the case with $z = 1$) now become either strictly  renormalizable or  supperrenormalizable, whereby  
the  strong  coupling problem is resolved.  For more details, we refer readers to \cite{LWWZ,BPSc} \footnote{ The mechanism requires
the coupling constants of the six-order derivative terms, represented by $\gamma_{n}\; (n = 5, 6, 7, 8)$ in the action (\ref{2.30}), must be very large $\gamma_{n} \simeq 1/\sqrt{\lambda - 1} \ge 10^7$.
This may introduce a new hierarchy \cite{KP13}, although it was argued that this is technically natural \cite{BPSb,BPSc}.}. 
 Note that in this case the strong coupling energy is given by \cite{BPSb,BPSc},
\bq
\lb{2.33}
\Lambda_{\omega} = \left|\lambda - 1\right|^{1/2} M_{pl},
\eq
instead of that given by Eq.(\ref{2.25}) for the HMT extension with a local U(1) symmetry.

With the non-projectability condition, static spacetimes with
$\beta_0  = 0$
have been extensively studied, see for example, \cite{LMP,CCOa,KS09,CCOb,Myung09,Mann09,CJ09,Park09}. Sine all of these 
works were carried out before the realization of the importance of the term $a_ia^i$ that could play in the IR stability \cite{BPSa,BPSb}, so  they all unfortunately belong to  
 the branch of the non-projectable  Ho\v{r}ava gravity  that is plagued with the instability problem \footnote{In \cite{Visser11}  attention has been called on the
self-consistency of the theory. In particular, one would like first  resolve   the instability problem before considering any application of the theory.}. 
Lately, the case
with   $\beta_0  \not= 0$ were studied in \cite{KK09,Park12,BS12a,BRS15}.  In particular, it was claimed that slowly rotating black holes do not exist in this extension \cite{BS12a}.
It was shown that this is incorrect \cite{Wang12,BS12b}, 
and slowly rotating black holes and  stars  indeed exist in all the four versions of Ho\v{r}ava gravity introduced in this review \cite{Wang13} \footnote{It should be noted that ``black holes" here 
are  defined by the existence of Killing/sound horizons.  But, particles can have velocities higher than that of light, once the Lorentz symmetry is broken \cite{LZbreaking}. Then,  Killing/sound horizons 
are no longer barriers to these particles, so such defined ``black holes'' are not really black to such particles.}. 

Cosmological perturbations were studied in \cite{KUYb,CHZb,FB12} and found that they are consistent with observations.  Cosmological constraints of Lorentz violation from
dark matter and dark energy were also investigated in \cite{BS11b,BIS,ABLS,ABILS}.  

In addition,  the odd terms given in Eq.(\ref{2.12})
violate the parity, which can polarize  primordial gravitational waves and  lead to direct observations \cite{TS09,WWZZ}. In particular,  
it was shown that, because of both parity violation and the nonadiabatic evolution of the modes due to a modified 
dispersion relationship, a large polarization of primordial gravitational waves becomes possible, and  could be  within the range of detection of the BB, TB and EB power spectra of the
 forthcoming CMB observations \cite{WWZZ}. Of course, this conclusion is not restricted to this version of Ho\v{r}ava gravity, and in principle it is true in all of these four versions presented
 in this review.

  \subsection{With  Non-projectability  $\&$  U(1) Symmetry}

A non-trivial generalization of the enlarged symmetry (\ref{2.20}) to the nonprojectable case $N = N(t, x)$ was realized   in
\cite{ZWWS,ZSWW,LMWZ},  and has been recently embedded into string theory by using the non-relativistic AdS/CFT correspondence \cite{JK1,JK2}. 
In addition, it was also found that it corresponds to the dynamical Newton-Cartan  geometry with twistless torsion (hypersurface orthogonal foliation) \cite{HO15}.
A precise dictionary was then constructed, which connect  all fields, their transformations and other
properties of the two corresponding theories.  Further, it was shown that  the U(1) symmetry  comes from the Bargmann extension of the local Galilean 
algebra that acts on the tangent space to the torsional Newton-Cartan geometries.  

The realization of the enlarged symmetry (\ref{2.20}) in the  nonprojectable case can be carried out by first noting the fact that ${\cal{N}}_{i}$ and $\sigma$ defined
in Eq.(\ref{2.25b})  are gauge-invariant under the local U(1) transformations and are a vector and scalar under Diff(M,{\cal{F}}), respectively. In addition, they
have dimensions two and four, respectively, 
\bqn
\lb{2.34}
\delta_{\alpha}{\cal{N}}_{i} = 0 = \delta_{\alpha}\sigma, \;\; [{\cal{N}}_i] = 2, \;\;
 [\sigma] = 4.
\eqn
Then, the quantity $\tilde{K}_{ij}$ defined by
\bqn
\lb{2.35}
\tilde{K}_{ij}&\equiv& \frac{1}{2N}\left(-\dot{g}_{ij} + \nabla_{i}{\cal{N}}_{j} + \nabla_{j}{\cal{N}}_{i}\right)\nb\\
&=& K_{ij}+\nabla_i\nabla_j\varphi+a_{(i}\nabla_{j)}\varphi,
\eqn
is also gauge-invariant under the  U(1) transformations. In addition, $\tilde{K}_{ij}$ has the same dimension as ${K}_{ij}$, i.e., $[\tilde{K}_{ij}] = 3$, from which one can see that
the quantity $ \tilde{\cal L}_K$,
\bq
\lb{2.36}
 \tilde{\cal L}_K \equiv \tilde{K}^{ij}\tilde{K}_{ij}-\lambda\tilde{K}^2,
 \eq
has dimension 6. In addition,     the quantity
\bq
\lb{2.37}
\tilde{\cal{L}}_{S} \equiv \sigma\left(Z_A+2\Lambda_g-R\right).
\eq
has  also dimension 6 and is gauge-invariant under the  U(1) transformations, where $Z_{A} \equiv \sigma_1 a^ia_i + \sigma_2 a^{i}_{\;\; i}$, with $\sigma_1$ and $\sigma_2$ 
being two dimensionless coupling constants. Then, the
action
\bqn
  \lb{2.38}
S_g&=&\zeta^2\int
dtd^3xN\sqrt{g}\big[\tilde{\cal L}_K-\gamma_1R-2\Lambda\nb\\
&&+\beta a^ia_i+\sigma\left(Z_A+2\Lambda_g-R\right)- {\cal
L}_{z>1}\big],
\eqn
is gauge-invariant under the  enlarged symmetry (\ref{2.20}), where $  {\cal L}_{z>1}$ denotes the potential part that contains spatial derivative operators higher than second-orders.
Inserting Eqs.(\ref{2.35})-(\ref{2.37}) into the above action, and then after integrating it partially, the total action takes the form \cite{LMWZ},
\bqn
\label{2.39}
S_g = \zeta^2 \int dtd^3x \sqrt{g} N \Big({\cal{L}}_{K}-{\cal{L}}_{V} + {\cal{L}}_{A}
+ {\cal{L}}_{\varphi} +{\cal{L}}_S\Big),\nb\\
\eqn
where ${\cal{L}}_{K}$ and ${\cal{L}}_{A}$ are given, respectively, by Eqs.(\ref{2.14}) and (\ref{2.24}), but now with 
 \bqn 
 \lb{2.40}
{\cal{L}}_{\varphi} &\equiv&  \varphi{\cal{G}}^{ij}\big(2K_{ij}+\nabla_i\nabla_j\varphi+a_i\nabla_j\varphi\big)\nb\\
& & +(1-\lambda)\Big[\big(\Delta\varphi+a_i\nabla^i\varphi\big)^2
+2\big(\Delta\varphi+a_i\nabla^i\varphi\big)K\Big]\nb\\
& & +\frac{1}{3}\hat{\cal G}^{ijlk}\Big[4\left(\nabla_{i}\nabla_{j}\varphi\right) a_{(k}\nabla_{l)}\varphi \nb\\
&&  ~~ + 5 \left(a_{(i}\nabla_{j)}\varphi\right) a_{(k}\nabla_{l)}\varphi + 2 \left(\nabla_{(i}\varphi\right)a_{j)(k}\nabla_{l)}\varphi \nb\\
&&~~ + 6K_{ij} a_{(l}\nabla_{k)}\varphi \Big],\nb\\
{\cal{L}}_S &\equiv&\sigma  (\sigma_1 a^ia_i+\sigma_2a^i_{\;\;i}),
 \eqn
where $\hat{\cal G}^{ijlk} \equiv  g^{il}g^{jk} - g^{ij}g^{kl}$, and ${\cal{L}}_{V}$ denotes the potential part of the theory made of $a_i,\; \nabla_i$ and $R_{ij}$ only. However,
as mentioned previously, once the projectability condition  is abandoned, it gives rise to a proliferation of a large number of independent coupling constants \cite{KP13,ZWWS}. 
To reduce the  number, one way is to  generalize the   Ho\v{r}ava detailed balance condition (\ref{2.16})  to
\cite{ZWWS},
\bq
\lb{2.41}
{\cal{L}}_{(V,D)} = \Big(E_{ij} \; A_{i}\Big)\left(\begin{matrix}
{\cal{G}}^{ijkl} & 0\cr 0 &  g^{ij}\cr
\end{matrix}
\right)
\left(\begin{matrix}
E_{kl}\cr A_{j}\cr
\end{matrix}
\right),
\eq
where    $E_{ij}$ and $A_{i}$ are given by
\bq
\lb{2/42}
E^{ij} \equiv \frac{1}{\sqrt{g}}\frac{\delta{W}_{g}}{\delta{g}_{ij}},\;\;\;
A^i \equiv \frac{1}{\sqrt{g}}\frac{\delta W_a}{\delta a_i}.
\eq
The super-potentials $W_g$ and $W_a$ are constructed as \footnote{Note that in \cite{VS} a different generalization was proposed, in which ${\cal{L}}_{(V,D)}$
takes the same form in terms of $E_{ij}$ and ${\cal{G}}^{ijkl}$, as that given by Eq.(\ref{2.16}), but with the superpotential including a term $a_ia^i$, i.e.,
$W_{g} = \frac{1}{w^{2}}\int_{\Sigma}{\omega_{3}(\Gamma)} + \int{d^3x \sqrt{g}\left[\mu(R - 2\Lambda) + \beta a_ia^i\right]}$. However, in this generalization, the six-order derivative term
$ \left(\Delta{a^{i}}\right)^{2}$ does not exist in the potential $ {\cal{L}}_{V}$, and is a particular case of Eq.(\ref{2.44}) with $\beta_8 = 0$. Without this term,
the sixth-order derivative terms are absent for scalar perturbations. As a result,  the corresponding  theory is not power-counting renormalizable  \cite{ZWWS}.},
\bqn
\lb{2.43}
W_{g}  &=& \frac{1}{w^{2}}\int_{\Sigma}{\omega_{3}(\Gamma)} + \mu \int{d^3x \sqrt{g}\Big(R - 2\Lambda\Big)},\nb\\
W_{a} &=& \int{d^{3}x \sqrt{g} \sum_{n= 0}^{1}{{\cal{B}}_{n}a^{i}\Delta^{n}{a_{i}}}},
\eqn
where  
${\cal{B}}_{n}$ are coupling  constants. However, to have a healthy infrared limit,   as mentioned previously, 
the detailed balance condition needs to be broken softly, by adding all the low dimensional operators, so that the potential finally takes the form
\cite{ZWWS},
\bqn
\lb{2.44}
 {\cal{L}}_{V} &=&  2\Lambda  -  \Big(\beta_0  a_{i}a^{i}- \gamma_1R\Big)
+ \frac{1}{\zeta^{2}} \Big(\gamma_{2}R^{2} +  \gamma_{3}  R_{ij}R^{ij}\Big)\nb\\
& & + \frac{1}{\zeta^{2}}\Bigg[\beta_{1} \left(a_{i}a^{i}\right)^{2} + \beta_{2} \left(a^{i}_{\;\;i}\right)^{2}
+ \beta_{3} \left(a_{i}a^{i}\right)a^{j}_{\;\;j} \nb\\
& & + \beta_{4} a^{ij}a_{ij} + \beta_{5}
\left(a_{i}a^{i}\right)R + \beta_{6} a_{i}a_{j}R^{ij} + \beta_{7} Ra^{i}_{\;\;i}\Bigg]\nb\\
& &
 +  \frac{1}{\zeta^{4}}\Bigg[\gamma_{5}C_{ij}C^{ij}  + \beta_{8} \left(\Delta{a^{i}}\right)^{2}\Bigg],
 \eqn
where all the coefficients, $ \beta_{n}$ and $\gamma_{n}$, are
dimensionless and arbitrary, except for the ones of the sixth-order derivative terms,  $\gamma_{5}$ and $\beta_{8}$, which are positive,
\bq
\lb{2.45}
 \gamma_{5} > 0, \quad \beta_{8} >  0, 
 \eq
 as can be seen from Eqs.(\ref{2.41})-(\ref{2.43}). 
The Cotton tensor $C_{ij}$  is defined as
\bq
\lb{2.46}
C^{ij} =  \frac{e^{ijk}}{\sqrt{g}} \nabla_{k}\Big(R^{j}_{l} - \frac{1}{4}R\delta^{j}_{l}\Big).
\eq
  In terms of $R_{ij}$ and $R$, we have \cite{ZWWS},
\bqn
\lb{2.47}
C_{ij}C^{ij}
&=& \frac{1}{2}R^{3} - \frac{5}{2}RR_{ij}R^{ij} + 3 R^{i}_{j}R^{j}_{k}R^{k}_{i}  +\frac{3}{8}R\Delta R\nb\\
& &  +
\left(\nabla_{i}R_{jk}\right) \left(\nabla^{i}R^{jk}\right) +   \nabla_{k} G^{k},
\eqn
where \footnote{Note that, from  Gauss's theorem,  the integral of $ \nabla_{k} G^{k}$ yields,
$\int{dt dx^3 N\sqrt{g} \nabla_k G^k}$ $=   - \int{dt dx^3 N\sqrt{g} a_k G^k} + \int_{\Sigma_t}{dt G^k dS_k}$. }
\lb{2.48}
\bqn
G^{k}\equiv \frac{1}{2} R^{jk} \nabla_j R - R_{ij} \nabla^j R^{ik}-\frac{3}{8}R\nabla^k R.
\eqn

When $\sigma_1 = \sigma_2 = 0$, it was shown that the spin-0 graviton is eliminated \cite{ZWWS,ZSWW,LMWZ}. This was further  confirmed by Hamiltonian analysis \cite{MSW}.
For the universal coupling with matter given by  Eqs.(\ref{2.25a}) - (\ref{2.25c}), all the PPN parameters were calculated and given explicitly in terms of the coupling constants \cite{LW13,LMWZ}. 
In particular, it was found that with the choice of the parameters $a_1$ and $a_2$ given by Eq.(\ref{2.25d}), the relativistic results (\ref{PPNa}) are also obtained.

When  $\sigma_1  \sigma_2 \not= 0$ the spin-0 graviton appears \cite{LMWZ,MSW}. It was shown that the theory is ghost-free   for the choice of $\lambda$ given by
Eq.(\ref{2.19c}), and the spin-0 graviton is stable   in both of the IR and UV, provided that the following holds \cite{LMWZ},
 \bq
 \lb{2.49}
\beta_8 > 0, \quad \sigma^{-}_{2} < \sigma_2 < \sigma^{+}_{2},
\eq
where $\sigma_{2}^{\pm} \equiv 4\left(-\gamma_1 \pm \sqrt{{\gamma}_{1}^2 - {\beta_0}/{2}}\right)$ and 
$ {\gamma}_{1}^2 - {\beta_0}/{2} \ge 0$.
In this case,  the analysis of the PPN parameters shows that the corresponding theory is consistent with all the 
solar system  tests in a large region of the phase space \cite{LW13,LMWZ}. In particular, when the two coupling constants $ \sigma_2$ and $\beta_0$ satisfy the relations 
\cite{LMWZ}
\bq 
\lb{2.49} 
\sigma_2 = 4(1-a_1),\;\;\; \beta_0 =
-2(\gamma_1+1), 
\eq 
 the relativistic  values  of the PPN parameters given by Eq.(\ref{PPNa}) can be still achieved.
 
The strong coupling problem appearing in the gravitational and/or matter sectors can be resolved \cite{ZSWW}, by the BPS mechanism introduced previously \cite{BPSc}. 

The consistency of the theory with cosmology was studied in \cite{ZHW} when matter is minimally coupled to the theory. For the universal coupling of Eqs.(\ref{2.25a}) -(\ref{2.25c}),
such studies have not been worked out, yet \cite{MWWZ}. 

The effects of parity violation on non-Gaussianities of primordial gravitational waves were also studied recently \cite{ZZHWW}. By calculating the three point function, it was found that the 
leading-order contributions
 to the non-Gaussianities come from the usual second-order derivative terms, which produce the same bispectrum as that found in GR. The contributions from high-order spatial n-th derivative 
 terms are always suppressed by a factor $(H/M_*)^{n-2}\; (n\ge 3)$, where $H$ denotes the inflationary energy and $M_*$ the suppression mass scale of the high-order spatial derivative operators 
 of the theory. Thus, the next leading-order contributions come from the 3-dimensional gravitational Chern-Simons term. With some reasonable arguments, it is shown that this 3-dimensional 
 operator is the only one that violates the parity and in the meantime has non-vanishing contributions to non-Gaussianities.
 
 Finally we note that, instead of introducing the U(1) gauge field $A$ and the pre-Newtonian potential $\varphi$, the authors in \cite{CKO} introduced two Lagrange multipliers $A_{CKO}$ 
 and  $\Lambda_{CKO}$   (Don't confuse with the U(1) gauge field $A$ and the cosmological  constant $\Lambda$ introduced above), and found that in the non-projectable case 
 the spin-0 graviton can be  eliminated for certain choices of the coupling   constants of the theory, while for other choices  the spin-0 graviton is still present. However, it is not clear which symmetry,
 if there is one,  will preserve these particular choices, so that the spin-0 graviton is always absent. Otherwise, it will appear generically once radiative corrections are taken into account. 
 It is interesting to note that, in contrast to the non-projectable case,   in the projectable case the spin-0 graviton is always eliminated  \cite{CKO}. 
 
  In addition,  Ho\v{r}ava gravity with mix spatial and time derivatives has been also studied recently  in order to avoid unacceptable violations of Lorentz invariance in the matter sector \cite{CGS1,CGS2}.
  Unfortunately, it was found that  the theory contains four propagating degrees of freedom, as opposed to three in the standard Ho\v{r}ava gravity, and the new degree of freedom is another scalar graviton,
  which is unstable at low energies \cite{CCGS}.  
  
  \subsection{Covariantization of  Ho\v{r}ava Gravity}
  
As GR is general covariant and consistent with all the observations, both cosmological and astrophysical, it is desired to write Ho\v{r}ava Gravity  
also in a general covariant form  \footnote{This process in general introduces new physics, for example, the appearance of
the instantaneous mode in the khronometric theory, which is absent in both Ho\v{r}ava and Einstein-aether theories, as shown below.}, 
in order to see more clearly  the difference between the two theories. One may take one step further and consider such a process as a mechanism 
to obtain   the IR limit of the UV complete theory of Ho\v{r}ava gravity with the symmetry of general covariance.  

To restore the full general covariance, one way is to make the foliation {\em dynamical} by using the St\"uckelberg trick \cite{RRA},  in which the
space-like hypersurfaces of $t = $ constant in the (3+1) ADM decompositions   are replaced  by a  scalar field $\phi = $ constant, which is always time-like \cite{BPS,GKS09}, 
\bq
\lb{2.50}
u_{\mu} u^{\mu} = -1, \;\; u_{\mu} \equiv \frac{\phi_{,\mu}}{\sqrt{X}}, \;\; X \equiv  - \gamma^{\mu\nu}\phi_{, \mu} \phi_{, \nu} > 0.
\eq
Clearly, choosing $\phi = t$, the original foliations are obtained. This gauge choice is often referred to as {\em the unitary gauge} \cite{BPS}. Then, the gauge symmetry  $ t \rightarrow  \xi_0(t)$ translates 
into the symmetry of the St\"uckelberg field,
\bq
\lb{2.51}
\phi \rightarrow \tilde{\phi} = f(\phi),  
\eq
where $ f(\phi)$ is an arbitrary monotonic function of $\phi$ only, so that 
\bq
\lb{2.52}
\tilde u_{\mu} \equiv \frac{\tilde\phi_{,\mu}}{\sqrt{\tilde{X}}} = \frac{ \epsilon_f\phi_{,\mu}}{\sqrt{X}} = \epsilon_f  u_{\mu},
\eq
where $\epsilon_f  \equiv {\mbox{Sign}}[df(\phi)/d\phi]$. Once $u_{\mu}$ is introduced, all other geometric quantities can be constructed from $\gamma_{\mu\nu}$ and $u_{\mu}$ by following Israel's 
method for space-like hypersurfaces \cite{Israel,WS08}. For example, the extrinsic curvature ${\cal{K}}_{\mu\nu}$ is given by,
\bqn
\lb{2.53}
{\cal{K}}_{\mu\nu} = h_{\lambda\mu}D^{\lambda}u_{\nu} =\frac{1}{\sqrt{X}}h^{\alpha}_{\mu}h^{\beta}_{\nu} D_{\alpha}D_{\beta}\phi,
\eqn
where  $D^{\lambda}$  denotes the covariant derivative with respect to $\gamma_{\mu\nu}$, and $h_{\mu\nu}$ is the projection operator, defined as 
\bq
\lb{2.53a}
h_{\mu\nu} \equiv \gamma_{\mu\nu} + u_{\mu}u_{\nu}.
\eq	

Defining the compatible covariant derivative $\nabla_{\mu}$ on the hypersurfaces $\phi = $ constant as,
\bq
\lb{2.54}
\nabla_{\mu} T^{\alpha_{1} ... \alpha_{k}}_{\beta_{1} ... \beta_{l}} \equiv h^{\alpha_{1}}_{\delta_{1}} ... h^{\alpha_{k}}_{\delta_{k}}
 h^{\sigma_{1}}_{\beta_{1}} ...  h^{\sigma_{l}}_{\beta_{l}} h^{\nu}_{\mu} D_{\nu} T^{\delta_{1} ... \delta_{k}}_{\sigma_1 ... \sigma_l},
\eq
we find that
\bqn
\lb{2.55}
{\cal{R}}_{\mu\nu\lambda\rho} &=& h^{\alpha}_{\mu} h^{\beta}_{\nu} h^{\gamma}_{\lambda} h^{\delta}_{\rho} \; {}^{(4)}R_{\alpha\beta\gamma\delta} +
{\cal{K}}_{\nu\lambda}{\cal{K}}_{\mu\rho} - {\cal{K}}_{\nu\rho}{\cal{K}}_{\mu\lambda},\nb\\
\nabla_{\mu} {\cal{R}}_{\nu\lambda} &=& h^{\alpha}_{\mu} h^{\beta}_{\nu} h^{\gamma}_{\lambda} D_{\alpha} {\cal{R}}_{\beta\gamma},
\eqn
where ${}^{(4)}R_{\alpha\beta\gamma\delta}$ denotes the 4D Riemann tensor made out of the  metric $\gamma_{\mu\nu}$, and ${\cal{R}}_{\mu\nu\lambda\rho}$ the intrinsic 
Riemann tensor with 
 ${\cal{R}}_{\mu\nu} \equiv {\cal{R}}^{\lambda}_{\;\;\mu\lambda\nu}$. Thus, we find that  
 \bqn
 \lb{2.56}
&&  {\cal{R}}_{\mu\nu} = h^{\alpha}_{\mu} h^{\beta}_{\nu} \; {}^{(4)}R_{\alpha\beta} + u^{\alpha}u^{\beta} \; {}^{(4)}R_{\alpha\mu\beta\nu} \nb\\
 && ~~~~~~~~~ +
 {\cal{K}}_{\mu\lambda}{\cal{K}}^{\lambda}_{\;\; \nu}  - {\cal{K}}{\cal{K}}_{\mu\nu},\nb\\
&&  {\cal{R}} \equiv h^{\mu\nu} {\cal{R}}_{\mu\nu} = {}^{(4)}R + {\cal{K}}^2 - {\cal{K}}_{\mu\nu} {\cal{K}}^{\mu\nu} + 2D_{\alpha}G^{\alpha},\nb\\
&&  {\cal{G}}^{\alpha} \equiv u^{\lambda}D_{\lambda}u^{\alpha} - u^{\alpha}D_{\lambda}u^{\lambda}.
 \eqn
Note that in writing the above expressions, we had used the relation \cite{Israel,WS08},
 \bq
 \lb{2.57}
{}^{(4)}R_{\alpha\beta}u^{\alpha}u^{\beta} =  {\cal{K}}^2 -  {\cal{K}}_{\alpha\beta}{\cal{K}}^{\alpha\beta}  + D_{\alpha} {\cal{G}}^{\alpha},
 \eq
where $ {\cal{K}} \equiv  {\cal{K}}^{\lambda}_{\lambda}$. Then, to obtain the relativistic version of Ho\v{r}ava gravity, one can simply identify the quantities appearing in 
Ho\v{r}ava gravity  in the ADM decompositions with those obtained in the unitary gauge, which leads to  the following replacements in the action,
 \bqn
 \lb{2.58}
N &\rightarrow& \frac{1}{\sqrt{X}},\nb\\
 \nabla_{i} &\rightarrow& \nabla_{\lambda},\nb\\
a_i &\rightarrow& a_{\lambda},\nb\\
\partial_{\perp}\varphi &\rightarrow& u^\nu h^\mu_\nu D_\mu \varphi,\nb\\
  R_{ij} &\rightarrow& {\cal{R}}_{\mu\nu},\nb\\
  K_{ij} &\rightarrow& {\cal{K}}_{\mu\nu},\nb\\
C^{ij} &\rightarrow& C^{\mu\nu},  
 \eqn
 where
 \bqn
 \lb{2.59}
&& \partial_{\perp} \equiv \frac{1}{N}\left(-\partial_{t} + N^i\nabla_{i}\right),\nb\\
 && C^{\mu\nu} \equiv \eta^{\mu\alpha\beta}\nabla_{\alpha}\left({\cal{R}}^{\nu}_{\beta} - \frac{1}{4} \delta^{\nu}_{\beta} {\cal{R}}\right),
 \eqn
where $\eta^{\mu\alpha\beta}$ is the three-dimensional volume element defined as $\eta^{\mu\alpha\beta} = \eta^{\mu\alpha\beta\delta}u_{\delta}$, and
$\eta^{\mu\alpha\beta\delta}$ denotes the  four-dimensional volume element with $\nabla_{\alpha} \eta^{\mu\nu\beta} = 0$ \cite{GKS09}.

In the case with an extra U(1)  symmetry, we also replace the gauge field $A$ and the pre-Newton potential $\varphi$, respectively, by \cite{LMW17},
\bqn
\lb{2.60}
A  &\rightarrow&  \frac{1}{\sqrt{X}}\left(\sigma + u^\nu h^\mu_\nu D_\mu \varphi  + \frac{1}{2}
 \left(\nabla_\lambda\varphi\right)^2\right),\nb\\
 \varphi & \rightarrow &   \varphi, 
\eqn
where $\sigma$ defined by Eq.(\ref{2.25b}) and $\varphi$  are scalars under both the U(1)   and the Diff($M, \; {\cal{F}}$)
transformations. Therefore, the resulting theories from the ones with the local U(1) symmetry will contain two Lagrangian multipliers $\sigma$ and $\varphi$, each of them acts like a scalar field, but none of
them is dynamical, in contrast to the St\"uckelberg field $\phi$, introduced above. This is similar to what was done in \cite{CKO}, and it would be very interesting to find out if they are related one to the other.

Applying the above St\"uckelberg trick to the healthy extension \cite{BPSa,BPSb}, it can be shown that the resulting action in the IR is identical to the hypersurface-orthogonal 
Einstein-aether theory \cite{JM01,Jacob07}, as shown
explicitly  in \cite{Jacob,Jacob13}, provided that  the aether four-vector $u_{\mu}$ satisfies the  hypersurface-orthogonal conditions,  
\bq
\lb{2.61}
u_{[\alpha}D_{\beta}u_{\lambda]} = 0,
\eq
which are necessary and sufficient conditions for    $u_{\mu}$ to be given by Eq.(\ref{2.50}) \cite{Wald94}. From Eq.(\ref{2.61}) it can be shown that
\bq
\lb{2.62}
\omega^2 \equiv  a^{\mu}a_{\mu} + \big(D_{\alpha}u_{\beta}\big)\big(D^{\alpha}u^{\beta}\big) -   \big(D_{\alpha}u_{\beta}\big)\big(D^{\beta}u^{\alpha}\big),
\eq
vanishes identically, where $a_{\mu} \equiv u^{\lambda}D_{\lambda}u_{\mu}$. Then, one can add the term,
\bq
\lb{2.63}
\Delta S_{\ae}   \equiv c_{0} \int{dx^4\sqrt{-\gamma} \;  \omega^2}, 
\eq
where $c_{0}$ is an arbitrary constant and $\gamma \equiv {\mbox{det}} (\gamma_{\mu\nu})$, into the action of the Einstein-aether theory \cite{JM01,Jacob07},
\bqn
\lb{2.64}
S_{\ae} &=&  \int d^{4}x \sqrt{-\gamma}\Big[c_1\left(D_{\mu}u_{\nu}\right)^2 + c_2 \left(D_{\mu}u^{\mu}\right)^2\nb\\
&& ~~  + c_3   \left(D^{\mu}u^{\nu}\right)\left( D_{\nu}u_{\mu}\right)   - c_4 a^{\mu}a_{\mu} \Big],
\eqn
to eliminate one of the four independent coupling constants $c_i\; (i = 1, 2, 3, 4)$ of  the Einstein-aether theory, so that there are only three independent coupling constants
in terms of the St\"uckelberg field $\phi$, which will be referred to as {\em the khronon field}, and the corresponding theory the {\em the khronometric  theory} \cite{BPSb,BS11},
which was also referred to as the ``T-theory'' in \cite{Jacob,Jacob13}. 

Some comments are in order. First, the action of Eq.(\ref{2.64}) is the most general action of the Einstein-aether theory in which the field equations for $\gamma_{\mu\nu}$ and $u_{\mu}$
are the second-order differential equations \cite{JM01,Jacob07}. In this theory, in addition to the spin-2 graviton encountered in GR, there are two extra modes, the spin-1 and spin-0 gravitons,
each of them moves with a different velocity $v_s$ \cite{JM01,Jacob07}. To avoid the gravitational \v{C}erenkov effects \cite{EMS}, each of them must be not less than the speed of light,
$v_s \ge c$. Second, the khronometric  theory is fundamentally different from the Einstein-aether theory even in the domain where the   hypersurface-orthogonal conditions (\ref{2.61}) 
hold. This can be seen clearly by the fact that there is an additional mode, the instantaneous propagation of signals, in the khronometric  theory \cite{BS11}, which is absent in the 
Einstein-aether gravity. This is closely related to the fact that in terms of the khronon field $\phi$, the action (\ref{2.64}) are in general fourth-order differential equations. It is the existence of this mode
that may provide a mechanism to restore the second law of thermodynamics \cite{BS11}. It is also interesting to note that this instantaneous mode does not exist in the healthy extension either
\cite{BPSa,BPSb}. It is introduced by the covariantization process with the use of  the St\"uckelberg trick. 

We also note that in \cite{BPS,BPSb}, the covariantization of the projectable Ho\v{r}ava gravity was also studied, and showed that this minimal theory is always fine-tuning due to the instability of the spin-0 graviton 
in this version of Ho\v{r}ava gravity.

 \section{Black Holes $\&$ Thermodynamics}
 \renewcommand{\theequation}{3.\arabic{equation}} \setcounter{equation}{0}

Black holes   have been   objects of intense study both theoretically and observationally over half a
century now \cite{BHs1,BHs2}, and so far there are many solid  observational evidences for their existence in our universe, especially after the first detection of gravitational waves from two binary
black holes on September 14,  2015 by aLIGO \cite{GW}. 
Theoretically,  such investigations   have been playing a fundamental   role in the understanding of the nature of  gravity in general, and QG in particular.
They  started with  the discovery of the laws of black hole mechanics \cite{BCH} and Hawking radiation \cite{HawkingR}, which  led to  the  profound
 recognition of the thermodynamic interpretation of the four laws \cite{Bekenstein73} and  the reconstruction of GR as the thermodynamic 
 limit of a more fundamental theory of gravity \cite{Jacobson95}.  More recently, they play the  essential role  in the understanding of  the AdS/CFT correspondence
  \cite{AdSCFTa,AdSCFTb,AdSCFTc,AdSCFTd,AdSCFTe} and firewalls \cite{BPZ,AMPS}.

\begin{figure}[tbp]
\centering
\includegraphics[width=8cm]{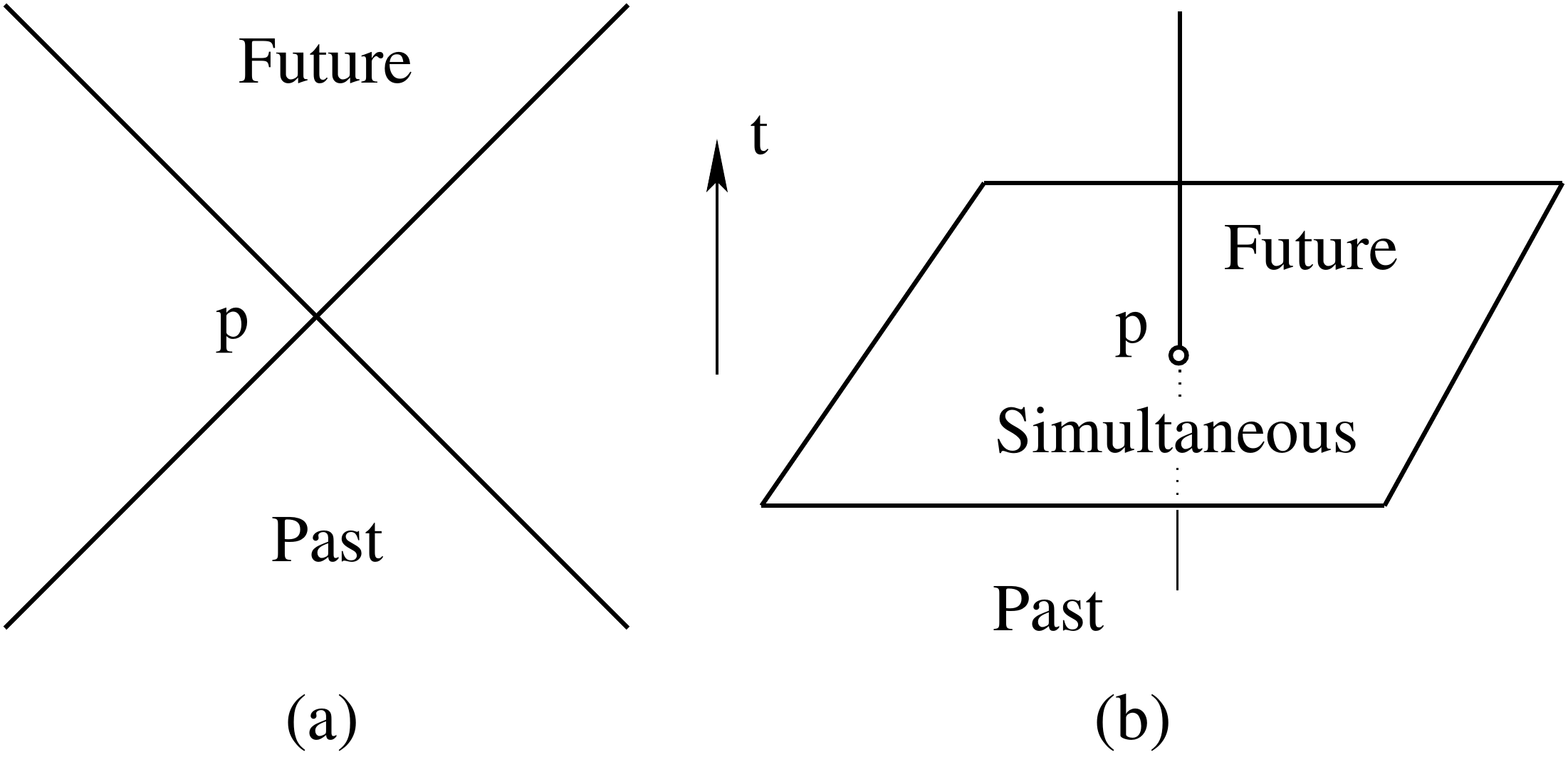}
\caption{ (a) The light cone of the event $p$ in special relativity. (b) The causal structure of the point $p$  in theories with broken LI. This figure is adopted from \cite{GLLSW}.}
\label{fig2}
\end{figure}

 Lately, such studies have gained  further momenta in the framework of gravitational theories with broken LI, including   Ho\v{r}ava gravity. In such theories, 
 due to the breaking of LI, the dispersion relation of a massive particle contains generically high-order momentum terms \cite{Muk,BC,TPS,WSV,Visser11,Padilla,Clifton,Hreview},
\bq
\lb{3.1}
E^{2} = m^{2} +  c_{p}^2p^{2}\left(1 + \sum^{2(z-1)}_{n=1}{a_{n}\left(\frac{p}{M_{*}}\right)^{n}}\right),
\eq
from which we can see that both of the group and phase velocities, $v_g \equiv E/p$ and $v_p \equiv dE/dp$, become unbounded as $p \rightarrow \infty$,
where $E$ and $p$ are the energy and momentum of the particle considered, and $c_p$ and $a_n$'s are coefficients, depending on the
species of the particle, while $M_{*}$ is the suppression energy  scale of the higher-dimensional operators.
As an immediate result,    the causal
structure of the spacetimes in such theories is quite different from that given in GR,  where the light cone at a given point $p$ plays a fundamental 
role in determining the causal relationship of $p$ to other events [cf. Fig. \ref{fig2}].  However, once LI is broken, the causal  structure will be dramatically changed. 
For example, in the Newtonian theory,  time is absolute and the speeds of signals are not limited. Then, the
causal structure of  a given point $p$ is uniquely determined by the time difference, $\Delta{t} \equiv t_{p} - t_{q}$, between the two events.  
 In particular, if $\Delta{t} > 0$, the event $q$ is to the past of $p$; if $\Delta{t} < 0$, it  is to the future; and if $\Delta{t} = 0$, the two events are
simultaneous.  In  Ho\v{r}ava gravity, a similar situation occurs.  This immediately makes
 the  definitions of black holes given in GR \cite{HE73}   invalid.  
 
\begin{figure}[tbp]
\centering
\includegraphics[width=1\columnwidth]{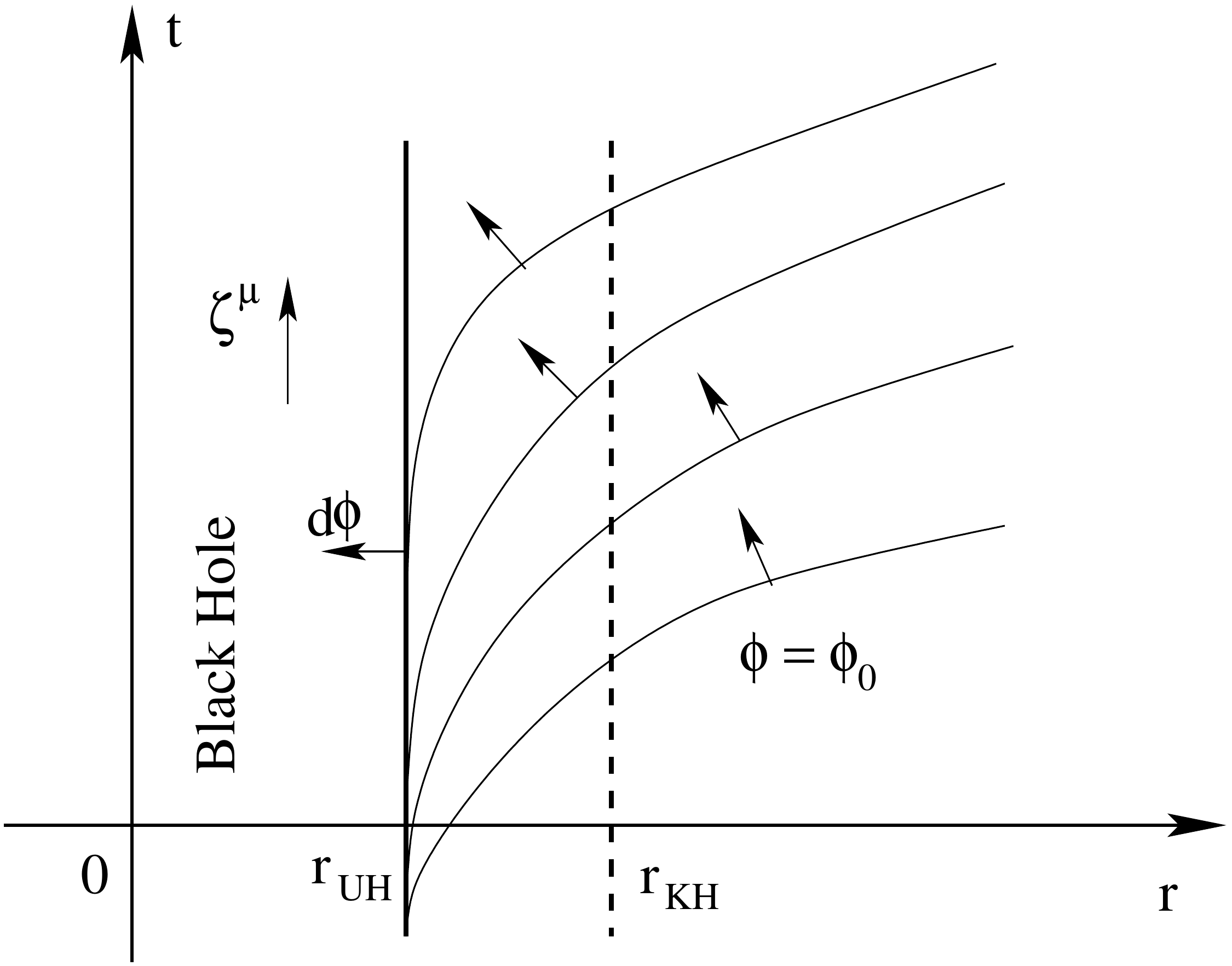}
\caption{Illustration of the bending of the $\phi$ = constant surfaces, and the existence of the universal horizon  in the Schwarzschild spacetime \cite{LACW}, 
where $\phi$ denotes the  globally timelike scalar  field, and $t$ and $r$ are the Painlev\'{e}-Gullstrand coordinates. Particles move always along the increasing direction of $\phi$. 
The Killing vector $\zeta^{\mu} = \delta^{\mu}_{t}$ always points upward at each point of the plane. The vertical dashed  line is   the location of the Killing horizon, 
$r=r_{KH}$. The universal horizon, denoted by the vertical solid  line, is  located at $r = r_{UH}$, which is always inside the Killing horizon.  This figure is adopted from \cite{LSW16}.} 
\label{Fig3}
\end{figure}

 To provide a proper definition of black holes,  anisotropic conformal boundaries \cite{HMT2} and  kinematics of particles  \cite{KM1,KM2,KM3,KM4,KK09,GLLSW} 
 have been studied   within the framework of Ho\v{r}ava gravity. Lately, a potential breakthrough  was the discovery that   there still exist absolute causal boundaries, the
so-called {\em universal horizons},  in  theories  with broken LI  \cite{BS11,BJS}.   Particles even with infinitely large velocities    would just move around on these
boundaries and  cannot escape to infinity.  The main idea is as follows. In a given spacetime, a globally timelike scalar field  $\phi$ might exist \cite{LACW} \footnote{In \cite{BS11,BJS}
the globally timelike scalar field is identified to the hypersurface-orthogonal aether field $u_{\mu}$, in which the aether is part of the gravitational field, the existence of which 
violates LI. However, to apply such a concept to other theories (without an aether field), a generalization is needed. In \cite{LACW}, the hypersurface-orthogonal aether field 
was promoted to a field that plays the same role as a Killing vector does in GR.}. 
Then, similar to the 
Newtonian theory, this field defines a global absolute time, and all particles are assumed to move along the increasing direction of the timelike scalar field, so the causality is 
well defined, similar to the Newtonian case [Cf. Fig.~\ref{fig2}]. In such a spacetime, there may exist a surface as shown in  Fig.~\ref{Fig3}, denoted by the vertical solid line,
located at $r = r_{UH}$. Given that all particles move along the increasing direction of the timelike scalar field, from Fig.~\ref{Fig3} it is clear that a particle must cross this surface and move inward, 
once it arrives at it, no matter how large of its velocity is. This is an one-way membrane, and particles even with infinitely large speed cannot escape from it, once they are 
inside it. So, it acts as an absolute horizon to all particles (with any speed),  which is often called {the universal horizon} \cite{BS11,BJS,LACW}.  At the universal horizon, we have
$dt\cdot d\phi = 0$, or equivalently,
\bq
\lb{3.2}
\zeta \cdot u   = 0,
\eq
where $\zeta (\equiv \partial_t)$ denotes the asymptotically timelike Killing vector, and $u (\equiv u_{\lambda}dx^{\lambda})$ is defined in terms of the globally timelike scalar field  $\phi$ via Eq.(\ref{2.50}). 
It is because of this    that  the globally timelike scalar field $\phi$ is also called the khronon field. But, there is a fundamental difference between the scalar field introduced here and the one introduced in Section II.E. In 
particular, the one introduced in Section II.E  is part of the gravitational field and represents the extra degrees of the  gravitational sector, while the one introduced here is a test field, which plays the same role as
a Killing vector field does, and no effects on the spacetime, but simply describes the properties of  such a given  spacetime.  So, whether such  a field exists or not only indicates whether a globally 
timelike scalar $\phi$ can be defined or not in such a given spacetime.

A Killing vector in a given spacetime background satisfies the Killing equations,  $D_{(\alpha}\zeta_{\beta)} = 0$. To define the globally  timelike scalar field  $\phi$, one way is to adopt the
equations for the khronon field given by the general action (\ref{2.64}), which can be written in the form \cite{Jacob13} \footnote{The quantity $\sigma$ introduced here
must not be confused with one 
introduced in Eq.(\ref{2.25b}).},
\bqn
\lb{3.3}
S_{\phi} &=&  \int d^{4}x \sqrt{-\gamma}\Bigg[\frac{1}{3}c_{\theta} \theta^2 + c_\sigma\sigma^2  - c_a a^2 + c_{\omega} \omega^2\Bigg], ~~~~
\eqn
where
\bqn
\lb{3.4a}
&& c_\theta \equiv c_{13}  + 3c_2, \quad
c_\sigma \equiv c_{13}, \nb\\
&& c_\omega \equiv c_1 - c_3, \quad 
c_a \equiv c_{14},\nb\\
\lb{3.4b}
&& D_{\beta}u_{\alpha}  = \frac{1}{3}\theta h_{\alpha\beta} + \sigma_{\alpha\beta} + \omega_{\alpha\beta} - a_\alpha u_\beta,
\eqn
with $c_{13} \equiv c_1 + c_3$, $c_{14} \equiv c_1 + c_4$, and
\bqn
\lb{3.5}
\theta &\equiv& D_{\lambda}u^{\lambda}, \quad h_{\alpha\beta} = \gamma_{\alpha\beta} + u_{\alpha} u_{\beta}, \nb\\
\sigma_{\alpha\beta} &\equiv & D_{(\beta}u_{\alpha)} + a_{(\alpha}u_{\beta)}  - \frac{1}{3}\theta h_{\alpha\beta},\nb\\
\omega_{\alpha\beta} &\equiv & D_{[\beta}u_{\alpha]} + a_{[\alpha}u_{\beta]},
\eqn
where $(A, B) \equiv (AB + BA)/2$. When $u_{\mu}$ is hypersurface-orthogonal, we have $\omega^2 =0$, so the last term in the above action vanishes
identically, as mentioned in Section II.E.  Even so, the  equation for $\phi$ still contains three free parameters, $(c_{\theta}, c_\sigma, c_a)$, and
their physical interpretations are not clear. 

Hence, the variation of  $S_{\phi}$ with respect to $\phi$ yields
the khronon equation,
\bqn
\lb{3.6}
D_{\mu} {\cal{A}}^{\mu}  = 0,
\eqn
where \cite{Wang12} \footnote{Notice the difference between the signatures of the metric chosen in this paper and the ones in \cite{Wang12}.},
\bqn
\lb{3.7}
{\cal{A}}^{\mu} &\equiv& \frac{\left(\delta^{\mu}_{\nu}  + u^{\mu}u_{\nu}\right)}{\sqrt{X}}\AE^{\nu},\nb\\
\AE^{\nu} &\equiv& D_{\gamma} J^{\gamma\nu} + c_4 a_{\gamma} D^{\nu}u^{\gamma},\nb\\
J^{\alpha}_{\;\;\;\mu} &\equiv&  \big(c_1g^{\alpha\beta}g_{\mu\nu} + c_2 \delta^{\alpha}_{\mu}\delta^{\beta}_{\nu}
+  c_3 \delta^{\alpha}_{\nu}\delta^{\beta}_{\mu}\nb\\
&&  ~~~ - c_4 u^{\alpha}u^{\beta} g_{\mu\nu}\big)D_{\beta}u^{\nu}.
\eqn

To solve the above equations for $\phi$, one can divide it into two steps: First solve Eq.(\ref{3.6}) in terms of $u_{\mu}$. Once $u_{\mu}$ is known, one can solve
$\phi$  from the definition,
\bq
\lb{3.8}
  {\phi_{,\mu}} = {\sqrt{ - \gamma^{\alpha\beta}\phi_{,\alpha} \phi_{,\beta}}} \; u_{\mu}.
\eq
Eq.(\ref{3.6}) is a second-order differential equation for $u_{\mu}$, and to uniquely determine it, two boundary conditions are needed.
These two conditions can be chosen as follows \cite{BS11}:
(i)  It will be aligned asymptotically with the timelike Killing vector,
\bq
\lb{3.8}
u^{\mu} \propto \zeta^{\mu}.
\eq
(ii) The second condition can be that the khronon has a regular future sound horizon, which
  is a null surface of the effective metric,
\bq
\lb{3.9}
\gamma^{(\phi)}_{\mu\nu} = \gamma_{\mu\nu} - \left(c_{\phi}^2 -1\right)u_{\mu}u_{\nu},
\eq
where $c_{\phi} (\equiv (c_{13}+c_2)/c_{14})$ denotes the speed of the khronon \footnote{To avoid the \v{C}erenkov effects,
 in the khronometric theory, or more general, in the Einstein-aether theory, $c_{\phi}$ is required to be no less than  the speed of light $c_{\phi} \ge c$ \cite{EMS}. However, since in the current
 case, the khronon is treated as a probe field, such requirement is not needed.}. 
 
 It is interesting to note that when $c_a = c_{14} = 0$, the khronon has an infinitely large velocity, and its sound
 horizon  now coincides with the universal horizon. Then, the second condition becomes to require that the universal horizon be regular. This is an interesting choice, and it also reduces
 the free parameters in the action (\ref{3.3}) from three, $(c_{\theta}, c_\sigma, c_a)$,  to two, $(c_{\theta}, c_\sigma)$. Out of these two only the ratio $c_{\theta}/c_\sigma$ appears in the 
 globally timelike field equation (\ref{3.6}). 
 
 In addition,  since $u_{\mu}$ is always timelike, Eq.(\ref{3.2}) is possible only inside a Killing horizon,  in which $\zeta$ becomes spacelike. That is, the radius of a universal horizon is always smaller than 
 that of a Killing horizon, which is physically expected, as the universal horizon represents the one-way membrane for particles with any large velocities, including the infinitely large one.

 Since they were first discovered in \cite{BS11,BJS}, universal horizons  have  been studied intensively, and already attracted lots of attention. These include universal horizons in static spacetimes \cite{Satheesh,UH1,UH2,UH3,UH4,UH5,UH6,SLWW,LSWW,UH7,UH8,UH9,UH10,UH11,UH13,UH14,UH15,UH16}, and the ones with rotation \cite{rUH1,rUH2,LSW16} \footnote{When in the process
 of  submitting   this review for publication, an interesting paper on Smarr Formula for Lorentz-breaking gravity \cite{PL17} appears.  The formula,  derived  by using the Noether charge analysis,
  is applicable not only to static cases
 but also with rotations.}.
  In particular, it was showed that  universal horizons  exist   
 in the three well-known black hole solutions:  Schwarzschild, Schwarzschild anti-de Sitter, and  Reissner-Nordstr\"om \cite{LGSW}, which are also solutions of  Ho\v{r}ava gravity \cite{GLLSW}.
 
 In addition,   universal horizons exist not only  in the low energy limit but also in the UV regime \cite{LSW16}.  
It is interesting to note   that in \cite{MM15},  the effects of higher-order derivative terms on the existence of universal horizons were studied, and found that, if a three-Ricci curvature squared term is 
joined in the ultraviolet modification,  the universal horizon appearing in the low energy limit was turned into  a spacelike singularity. This is possible, as the universal horizons might not be 
stable against nonlinear perturbations \cite{BS11}.
 
At the universal horizon, the first law of black hole mechanics exists for the neutral Einstein-aether black holes  \cite{BBMa}, provided that the surface gravity is defined by \cite{CLMV}, 
\bqn
 \lb{3.10}
\kappa_{UH} \equiv  \frac{1}{2} u^{\alpha} D_{\alpha} \left(u_{\lambda} \zeta^{\lambda}\right),
 \eqn
which was obtained by considering the peeling behavior of ray trajectories of constant khronon field $\phi$. However, for the charged Einstein-aether black holes,  such a first law is still absent  \cite{DWW}. 

Using the tunneling method,  Hawking radiation at the universal horizon for a scalar field that violates the local LI was studied, and found that the universal horizon radiates as a 
blackbody at a fixed temperature  \cite{BBMb}. A different approach was taken in  \cite{CLMV}, in which ray trajectories in such black hole backgrounds were studied, and evidence was found, 
which shows that Hawking radiation is associated with the universal horizon, while the ``lingering" of low-energy ray trajectories near the Killing horizon hints a reprocessing there. However, 
the study of a collapsing null shell showed that the mode passing across the shell is adiabatic at late time  \cite{MP15}. This implies that large black holes emit a thermal flux with a temperature fixed 
by the surface gravity of the Killing horizon. This, in turn, suggests that the universal horizon should play no role in the thermodynamic properties of these black holes, although it must be noted that 
in such a setting, the khronon field is not continuous across the collapsing null shell. As mentioned above,  a  globally-defined khronon plays an essential role in the causality of the theory,
so it is not clear how the results presented in \cite{MP15} will be changed once the continuity of the khronon field is imposed.  

On the other hand, using the Hamilton-Jacobi  method, quantum tunneling of both relativistic and non-relativistic particles at Killing as well as  universal horizons of Einstein-Maxwell-aether black holes
were studied \cite{DWWZ}, after higher-order curvature corrections are taken into account.  It was found  that only relativistic particles are created at the Killing horizon, and the corresponding radiation 
is thermal with a temperature exactly the same as that found in GR. In contrary, only non-relativistic particles are created at the universal horizon and are radiated out to infinity with a thermal spectrum. 
However, different species of particles, in general, experience different temperatures,
\bq
\lb{3.11}
T_{UH}^{z\ge 2} = \frac{2(z-1)}{z}\left(\frac{\kappa_{UH}}{2\pi}\right),
\eq
where $\kappa_{UH}$ is the surface gravity calculated from Eq.(\ref{3.10}) and $z$ is the exponent of the dominant term in the UV [cf. Eq.(\ref{3.1})].  When  $z = 2$ we have
the standard result, 
\bq
\lb{3.12}
T_{UH}^{z = 2} = \frac{\kappa_{UH}}{2\pi},
\eq
 which was first obtained in \cite{BBMb,CLMV}. 

Recently,  more careful studies of ray trajectories showed that the surface gravity for particles with a non-relativistic dispersion relation  (\ref{3.1}) is given by \cite{DL16},
\bq
\lb{3.13}
\kappa_{UH}^{z\ge 2} = \left(\frac{2(z-1)}{z}\right) \kappa_{UH},
\eq
so that Eq.(\ref{3.10}) is true only for particles with $z =2$. The same results were also obtained in \cite{Cropp16}. It is remarkable to note that in terms of $\kappa_{UH}^{z\ge 2}$
and $T_{UH}^{z\ge 2}$, the standard relationship between the temperature and surface gravity of a black hole still holds here,
\bq
\lb{3.14}
T_{UH}^{z \ge 2} = \frac{\kappa_{UH}^{z\ge 2}}{2\pi}.
\eq
Is this a coincidence? Without a deeper understanding of  thermodynamics of universal horizons, it is difficult to say. But, whenever case like this raises, it is worthwhile of paying some
special attention on it. In particular, does entropy of a universal horizon also depend on the dispersion relations of particles? 

To understand the problem better, another important issue is: Can universal horizons be formed from gravitational collapse? so they can naturally exist in our universe \cite{GLSW}. To
answer this question, the collapse of a spherically symmetric thin-shell in a flat background that finally forms a Schwarzschild black hole was studied \cite{SAM14}, in which the globally 
timelike scalar field is taken as the Cuscuton field \cite{ACG}, a scalar field with infinitely large sound speed. It was  shown that  an observer inside the universal horizon of the 
Schwarzschild radius cannot send a signal outside, after a stage in collapse, even using signals that propagate infinitely fast in the preferred frame. Then, it was argued   
that this universal horizon should be considered as the future boundary of the classical space-time. Lately, such studies were generalized to the Reissner-Nordstrom and Kerr
black holes \cite{MSA16} \footnote{It should be noted that in these studies the Schwarzschild-like coordinates were used, in which the metric outside of the collapsing thin shell is 
valid only for radius of the shell greater than the radii of the Killing horizons. But, universal horizons occur always inside  Killing horizons.   In addition, the Cuscuton theory in general 
does not have the symmetry (\ref{2.51}) that reflects the gauge symmetry  
$ t \rightarrow  \xi_0(t)$ of Ho\v{r}va gravity. For other related aspects of Cuscuton theory to  Ho\v{r}ava gravity and Einstein-aether theory, see \cite{BCCGS}.}. 

In addition, gravitational collapse of a scalar field in the Einstein-aether theory was studied in \cite{GEJ}, prior to the discovery of universal horizons. Lately, 
it was  revisited \cite{BCCS16}. However, due to the specially slicing of the spacetime carried out in \cite{GEJ}, the numerical simulations 
cannot penetrate  inside universal horizons, so it is not clear whether or not universal horizons have been formed in such simulations. 

In any case, the universal horizons defined by Eq.(\ref{3.2}) are in terms of the timelike Killing vector, which exists only in stationary spacetimes. To study the dynamical formation of 
universal horizons from gravitational collapse, a generalization of Eq.(\ref{3.2}) to non-stationary spacetimes is needed. One way, as first proposed in \cite{TWSW},  is to replace the 
timelike Killing vector by a Kodama-like vector \cite{Kodama80}, which reduces to the  timelike Killing vector in the stationary limit. To be more specific,  let us consider
spacetimes described by the metric, 
\bq
\lb{3.15}
ds^2 = g_{ab}dx^adx^b + {R}^2\left(x^0, x^1\right)d\Sigma_k^2, \; (a, b = 0, 1),
\eq
in the coordinates, $x^{\mu} = \left(x^0, x^1, \theta, \varpi\right), \; (\mu = 0, 1, 2, 3)$,  where $ k = 0, \pm 1$, and
 \bqn
\lb{3.16}
 d\Sigma^2_k =
  \begin{cases}
    d\theta^2+\sin^2\theta d\varpi^2,  &$k = 1$, \cr
    d\theta^2+d\varpi^2,                & $k = 0$, \cr
    d\theta^2+\sinh^2\theta d\varpi^2, & $k = -1$. \cr
  \end{cases}
 \eqn
The normal vector $n_{\mu}$ to the hypersurface $ {R} = C_0$ is given by,
\bq
\lb{3.17}
n_{\mu} \equiv \frac{\partial({R} -  C_0)}{\partial x^{\mu}} = \delta^{0}_{\mu} {R}_{0} + \delta^{1}_{\mu} {R}_{1},
\eq
where  $C_0$ is a constant and  ${R}_{a} \equiv \partial{R} /\partial x^a$. Setting
\bq
\lb{3.18}
\zeta^{\mu} \equiv \delta_{0}^{\mu} {R}_{1} - \delta_{1}^{\mu} {R}_{0},
\eq
we can see that $\zeta^{\mu}$ is always orthogonal to $n_{\mu}$,
$\zeta \cdot n = 0$.  
The vector $\zeta^{\mu}$ is often called Kodama vector and plays important roles in black hole thermodaynamics~\cite{Hayward:1998ee,Mukohyama:1999sp}. 
For spacetimes that are asymptotically flat there always exists a region, say, ${R} > {R}_{\infty}$,
in which $n_{\mu}$ and $\zeta^{\mu} $ are, respectively, space- and time-like, that is,
\bq
\lb{3.19}
\left. \left(n\cdot n\right) \right|_{{R} > {R}_{\infty}} > 0, \quad \left. \left(\zeta \cdot  \zeta\right)\right|_{{R} > {R}_{\infty}} < 0.
\eq
An apparent horizon may form at ${R}_{AH}$, at which
$n_{\mu} $ becomes null,
\bq
\lb{3.20}
\left. \left(n\cdot n\right) \right|_{{R} = {R}_{AH}} = 0,
\eq
where $ {R}_{AH} <  {R}_{\infty} $. Then, in the internal region $ {R} < {R}_{AH}$,  the normal vector $n_{\mu} $ becomes timelike.
Therefore, we have
\bq
\lb{3.21}
\left(n\cdot n\right)  = \begin{cases}
> 0, &  {R} > {R}_{AH}, \cr
= 0, & {R} = {R}_{AH}, \cr
< 0, & {R} < {R}_{AH}. \cr
\end{cases}
\eq
Since Eq.(\ref{3.19}) always holds, we must have
\bq
\lb{3.22}
\left(\zeta \cdot  \zeta\right)  =
\begin{cases}
 < 0, & {R} > {R}_{AH}, \cr
= 0, & {R} = {R}_{AH}, \cr
> 0, & {R} < {R}_{AH}, \cr
\end{cases}
\eq
that is, $\zeta^{\mu}$  becomes null on the apparent horizon, and spacelike inside it.

We define an {\em apparent universal horizon} as the hypersurface at which
\bq
\lb{3.23}
\left.\left(u \cdot \zeta\right)\right|_{{R} = {R}_{UH}} = 0.
\eq
Since $u_{\mu}$ is  globally timelike, Eq.(\ref{2.16}) is possible only when $\zeta_{\mu}$ is spacelike. Clearly, this is possible only  inside the
apparent horizon, that is, $  {R}_{UH} <  {R}_{AH}$.

In the static case, the apparent horizons defined above reduce to the Killing horizons, and the apparent universal horizons defined by Eq.(\ref{3.23})
are identical to those given by Eq.(\ref{3.2}).

The above definition is only for spacetimes in which the metric can be cast in the form (\ref{3.15}). However, the generalization to other cases is straightforward. In particular, 
it was generalized in terms of an optical scalar built with the preferred flow defined by the preferred spacetime foliation \cite{Maciel16}.

 \section{Non-relativistic Gauge/Gravity Duality}
  \renewcommand{\theequation}{4.\arabic{equation}} \setcounter{equation}{0}

 Anisotropic scaling plays a fundamental role in quantum phase transitions in condensed matter and ultracold atomic gases \cite{Cardy02,Sachdev13}.
Recently, such studies have  gained  considerable  momenta from the community of string theory in the content  of
gauge/gravity duality \cite{AGMOO,Maldacena11,Polchinski10}. This  is a duality between a  QFT in D-dimensions and a QG, such as string theory,
in (D+1)-dimensions.  An initial example was found  between the supersymmetric Yang-Mills gauge theory with maximal
supersymmetry in four-dimensions and a string theory on a five-dimensional anti-de Sitter space-time
in the low energy limit \cite{AdSCFTc}.  Soon, it was discovered that such a  duality is not restricted to the above systems, and can
be valid  among various theories and in different spacetime backgrounds \cite{AGMOO,Maldacena11,Polchinski10}.

One of the remarkable features of the duality is that it relates a strong coupling QFT to a weak coupling gravitational
theory, or vice versa. This is particularly attractive to condensed matter physicists, as it may provide hopes to understand  strong coupling systems encountered in
quantum phase transitions, by simply studying  the dual weakly  coupling gravitational theory \cite{Hartnoll09,McGreevy10,Horowitz10,Sachdev12,Cai15}. Otherwise, it has been found extremely difficult  to
study those systems. Such studies were  initiated in \cite{KLM}, in which it was shown that nonrelativistic QFTs that describe multicritical points in
certain magnetic materials and liquid crystals may be dual to  certain nonrelativistic  gravitational theories in the Lifshitz space-time background,
\bq
\lb{4.1}
ds^2 = - \left(\frac{r}{\ell}\right)^{2z} dt^2 + \left(\frac{r}{\ell}\right)^{2}dx^i dx^i + \left(\frac{\ell}{r}\right)^{2} dr^2,
\eq
where $z$ is a dynamical critical exponent, and $\ell$ a dimensional constant. Clearly, the above metric is invariant under the anisotropic scaling,
\bq
\lb{4.2}
 t \rightarrow b^{z} t, \quad x^i \rightarrow b x^i, \quad r  \rightarrow b^{-1}r.
\eq
  Thus, for $z \not= 1$ the relativistic scaling is broken in the sector ($t, x^i)$, and to have the above Lifshitz space-time as a solution of
GR, it is necessary to introduce gauge fields to create a preferred direction, so that the anisotropic scaling (\ref{4.2}) becomes possible.
In \cite{KLM}, this was realized by two p-form gauge fields with $p = 1, 2$, and was soon generalized to different  cases   \cite{Taylor16}.

It should be noted that the Lifshitz space-time  is singular at $r = 0$ \cite{KLM}, and this singularity is generic in the sense that it cannot be eliminated by
simply embedding  it to high-dimensional spacetimes, and that test particles/strings become infinitely excited when passing through the singularity \cite{CM11,HW13}.
To resolve this issue, various scenarios have been proposed. There have been also attempts to cover the singularity by horizons,
and replace  it by Lifshitz solitons \cite{Taylor16}.

On the other hand,   since the anisotropic scaling is built in by construction in Ho\v{r}ava gravity, it is natural to expect that the Ho\v{r}ava theory should provide a minimal holographic dual for
non-relativistic Lifshitz-type field theories with the anisotropic scaling and dynamical exponent $z$. Indeed, recently it was  showed that the Lifshitz spacetime (\ref{4.2})
is a vacuum solution of Ho\v{r}ava gravity in (2+1) dimensions, and that   the full structure of the $z=2$ anisotropic Weyl anomaly can be reproduced  in dual field theories \cite{GHMTc},
while  its minimal relativistic gravity counterpart yields only one of two independent central charges in the anomaly. 

Note that in \cite{GHMTc}, only the IR limit of the (2+1)-dimensional Ho\v{r}ava theory  was considered. Later,  the effects of high-order operators on 
the non-relativistic Lifshitz holography were studied \cite{WYTWDC},   and found that the Lifshitz space-time is still a solution of the full theory of the Ho\v{r}ava gravity. 
The effects of the high-oder operators on the space-time itself is simply to shift the Lifshitz dynamical exponent $z$. However,  the asymptotic behavior of a (probe) 
scalar field near the boundary gets dramatically modified in the UV limit, because of the presence of the high-order operators in this regime. Then, according to the gauge/gravity duality, 
this in turn affects the two-point correlation functions.

The above studies in the framework of Ho\v{r}ava gravity were soon generalized to various cases, in which various Lifshitz soliton, Lifshitz spacetimes with hyperscaling violations,
 and Lifshitz charged black hole solutions in terms of universal horizons \cite{LACW,UH4,UH5,UH6,SLWW,LSWW,UH7,UH11,UH14,UH16} as well as in terms of Killing horizons \cite{AY14,LKS16a,LKS16b} 
 were obtained and studied.
  
Recently, another important discovery of the non-relativistic gauge/gravity duality is the one-to-one correspondence between the Ho\v{r}ava gravity with the enlarged symmetry (\ref{2.20})
and the Newton-Cartan geometry (NCG) \cite{HO15}.  In particular, the projectable Ho\v{r}ava gravity discussed in Section II.B corresponds to the dynamical NCG without torsion, while
the  non-projectable  Ho\v{r}ava gravity discussed in Section II.D corresponds to the dynamical NCG with twistless torsion.  A precise dictionary between these two theories was established. 
Restricted to  (2+1) dimensions, the effective action for dynamical twistless torsional NCG  with $1<z\le 2$ was constructed by using the NCG invariance, and 
demonstrated that this exactly agrees with the most general forms of the  Ho\v{r}ava actions constructed in \cite{ZWWS,ZSWW,LMWZ}. Further,   the origin of the local U(1) symmetry  was identified 
 as coming from the Bargmann extension of the local Galilean algebra that acts on the tangent space to the torsional NCG.  Such studies have already attracted lots of attention and generalized 
 to other cases \cite{Taylor16}. 
 
 In addition, it was shown that the non-projectable Ho\v{r}ava gravity with the enlarged symmetry (\ref{2.20}) \cite{ZWWS,ZSWW,LMWZ}  can be also deduced from non-relativistic QFTs with conserved 
 particle number, by using  the  symmetries: time dependent spatial diffeomorphisms acting on the background metric and U(1) invariance acting on the background fields which couple to 
 particle number \cite{JK1,JK2}.   As Ho\v{r}ava gravity is presumed to be UV complete, in principle this duality  allows holography to move beyond the large $N$ limit  
 \cite{AdSCFTb,AdSCFTc,AdSCFTd,AdSCFTe}. 
 
 Finally, we would like to note that one may flip the logics around and study QG by using the gauge/gravity duality from well-known QFTs \cite{EH13,EH16}.

 \section {Quantization of Ho\v{r}ava Gravity}
  \renewcommand{\theequation}{5.\arabic{equation}} \setcounter{equation}{0}

 Despite numerous efforts and the vast literature on  Ho\v{r}ava gravity, so far its quantization has not been worked out in its general form, yet. Nevertheless, particular situations have
 been investigated, and some (promising)  results have been obtained. In particular, in (3+1)-dimensional spacetimes with the projectability and detailed balance conditions (\ref{2.15})-(\ref{2.18}),
 the renormalizability of Ho\v{r}ava gravity was shown to reduce to the one of the corresponding (2+1)-dimensional topologically massive gravity \cite{OR09}, by using stochastic quantization 
 \cite{PS81,DH87,Namiki91}. Even though the renormalizability of the latter has not been rigorously proven, it is often thought that it should be the case \cite{DY90}. As already pointed out in 
 \cite{OR09}, the equivalence of the renormalizability between the two theories is closely related by the detailed balance condition (\ref{2.16})-(\ref{2.18}), and it is not clear how to generalize it to 
 the case without this condition.  
 
 Lately, the above obstacle is circulated by properly  choosing a gauge that ensures the correct anisotropic scaling of the propagators and their uniform falloff at large frequencies 
 and momenta \cite{BBHSS}. It is this choice that  guarantees the counterterms required to absorb the loop divergences to be local and marginal or relevant with respect to the anisotropic scaling. 
 Then, gauge invariance of the counterterms is achieved by making use of the background-covariant formalism. 
 
 Along a similar line, Li { \em et al.}  studied the quantization of Ho\v{r}ava gravity both with and without the projectability condition (\ref{2.15}) in (1+1)-dimensional (2D) spacetimes \cite{Li14,Li16}. 
 In such spacetimes,  Einstein's theory  is trivial \cite{Jackiw85,Brown88,GKV02}. But, this is not the case of  Ho\v{r}ava gravity,  due to the   foliation-preserving diffeomorphism (\ref{2.3}) \cite{SVW11,EJ06},
 although the total degree of freedom of the theories is zero \cite{Li14,Li16}. In particular, in the projectable case, when only gravity is present, the system can be quantized by following the canonical 
 Dirac quantization \cite{Dirac64}, and the corresponding wavefunction is normalizable  \cite{Li14}.  It is remarkable to note that in this case the corresponding Hamilton can be written 
 in terms of a simple harmonic oscillator, whereby the quantization can be carried out quantum mechanically in the standard way.  When   minimally coupled to a scalar field, the momentum constraint can
 be solved explicitly in the case where the fundamental variables are functions of time only. In this case, the coupled system can also be quantized by following the Dirac process, and the corresponding 
 wavefunction is also normalizable.  A remarkable feature is that orderings of the operators from a classical Hamilton to a quantum mechanical one play a 
 fundamental role in order for the Wheeler-DeWitt equation to have nontrivial solutions. In addition, the space-time is well quantized, even when it is classically singular.

 In the non-projectable   case, the analysis of the 2D Hamiltonian structure shows that there are two first-class and two second-class constraints \cite{Li16}. Then, following Dirac one can quantize the theory by 
 first requiring that the two second-class constraints be strongly equal to zero, which can be carried out by replacing the Poisson bracket by the Dirac bracket  \cite{Dirac64}. The two first-class constraints give rise to the 
 Wheeler-DeWitt equations. It was found that in this case  the characteristics of classical spacetimes are encoded solely in the phase of the   solutions  of these equations.

 On the other hand, in CDTs \cite{CDTs} a preferred frame is also part of the process of quantization, which is quite similar to the foliation specified  in  Ho\v{r}ava gravity.  It was precisely this similarity that led 
 the authors of \cite{AGSW} to show the exact equivalence between the 2D CDT and the 2D projectable Ho\v{r}ava gravity. Such studies were further generalized to the case coupled with a scalar field \cite{AGJZ}
 and (2+1)-dimensional spacetimes \cite{ACCHKZ,BH15}. In particular,  in \cite{ACCHKZ} it was shown the existence of known and novel macroscopic phases of spacetime geometry, and found evidence for the consistency of these 
 phases with solutions to the equations of motion of classical Ho\v{r}ava gravity. In particular, the phase diagram seemingly contains a phase transition between a time-dependent de Sitter-like phase and a time-independent 
 phase.   In \cite{BH15},  spacetime condensation phenomena were considered and shown that a successful condensation in (2+1)-dimensions  can be obtained from a minisuperspace model of Ho\v{r}ava gravity, but not from
 GR.
 
 The close relations between CDTs and Ho\v{r}ava gravity have been further explored from the point of view of spectral dimensions. In particular, after  extending the definition of spectral dimension in
 fractal and lattice geometries  to theories on smooth spacetimes with anisotropic scaling,  Ho\v{r}ava showed that   a (d+1)-dimensional  spacetime with a dynamical critical exponent $z$ has the spectral dimension \cite{Horava09},
 \bq
 \lb{5.1}
 d_s=1+\frac{d}{z}.
 \eq
Thus, in the IR ($z=1$) the spectral dimension is identical to the macroscopic dimension $N (\equiv d + 1)$ of spacetimes, $N = d_s$.  But, in the UV ($z = d$) the  spectral dimension is always two, no matter what
dimensions of the spacetimes are in the IR. Therefore, (spectral) dimension is emergent and  it flows from two at short distances  to (d+1) at large ones. This was further confirmed in \cite{SVW_PRL,SVW_PRD}.  
Remarkably, this is also the qualitative behavior found  in CDTs 
\cite{AJL05} \footnote{Scale dependence of spacetime dimension was first considered in anisotropic cosmology \cite{HO86} and later in string theory where a thermodynamic dimension was also found to be two
at high temperature  \cite{AW88}.}. Even more surprising, the same results were  obtained in other theories of gravity \footnote{It should be noted that different definitions of dimension have been
adopted in these theories, and in principle they do not necessarily agree.}, such as asymptotic safety \cite{PP04,RS11}, LQG \cite{Modesto09,COT15}, spin foams \cite{MPM09}, 
noncommutative theory \cite{NHG}, Wheeler-DeWitt equation of GR \cite{Carlip09,Carlip12}, and causal set theory \cite{BBMM}, to name only a few of them. For more details, we refer readers to \cite{Carlip16}. 

In addition, it was shown recently that holographic renormalization of relativistic gravity in asymptotically Lifshitz spacetimes naturally reproduces the structure of Ho\v{r}ava gravity \cite{GHMTd}: 
The holographic counterterms induced near anisotropic infinity take the form of the action with the  anisotropic scaling (\ref{2.1}). 
 In the particular case of ($3+1$) bulk dimensions and $z=2$ asymptotic scaling near infinity,   a logarithmic counterterm was found, related to anisotropic Weyl anomaly of the dual conformal field theory.
  It is this counterterm that  reproduces precisely the action of conformal gravity at a $z=2$ Lifshitz point in ($2+1$) dimensions, which is of anisotropic local Weyl invariance and satisfies the 
detailed balance condition. It was also shown how the detailed balance is a consequence of relations among holographic counterterms. A similar relation also holds in the relativistic case of 
holography in AdS$_{5}$. Upon analytic continuation, similar to the relativistic case \cite{Maldacena11,Maldacena03,HS11}, the action of anisotropic conformal gravity produces the square of the wavefunction of the dual system.

 It is also very interesting to note that  recently one-loop effective action and beta functions of the projectable Ho\v{r}ava gravity were derived by using  the heat-kernel coefficients for Laplacian operators 
 obeying anisotropic dispersion relations in curved spacetimes  \cite{OSS14,OSS15}, and found that  the Gaussian fixed point  is  an infrared attractor for the renormalization group flow of Newton's constant, 
 and the high-energy phase of the theory is screened by a 
 Landau pole. When  coupled to $n$ Lifshitz scalars,  the   Gaussian fixed point ensures that the theory is asymptotically free in the large-N expansion, indicating that the theory  is perturbatively renormalizable. 
 Earlier  studies along the same line were carried out in \cite{KP13,OneLoop1,OneLoop2,OneLoop3,OneLoop4,OneLoop5,OneLoop6,OneLoop7}, and various  results were obtained.

 \section{Concluding Remarks}
 
 In this review, we have summarized  the recent developments in Ho\v{r}ava gravity. In particular, after giving a brief introduction of the general ideas of  Ho\v{r}ava at the beginning of Section II, we have pointed  out some 
 potential issues presented in the original incarnation of the theory, and then introduced four most-studied modifications, depending on the facts being with or without the projectability condition
 and a local U(1) symmetry. In the presentation for each of these four versions, we have stated clearly their  current status in terms of self-consistency of the theory, consistency with experiments, mainly with solar system tests and 
 cosmological observations.  
 
 In the projectable case without the local U(1) symmetry presented in Section II.A, {\em the minimal theory}, the main issues are
 instability \cite{SVWb,WM10,BS}, and strong coupling of the spin-0 graviton \cite{KA,WW10}. The perturbation analysis  shows that  the Minkowski spacetime is not stable  due to the presence of the spin-0 graviton 
 \cite{SVWb,WM10,BS}, which questions the viability of the theory (although  the de  Sitter spacetime is  \cite{WW10,HWW}).  To solve the strong coupling problem, one way is to borrow the well-known Vainshtein 
 mechanism \cite{Vainshtein:1972sx}. This has been done in the static  and cosmological settings \cite{Muk,Izumi:2011eh,GMW}, but for more general cases it has not been worked out, yet. 
   
  In the projectable case with the local U(1) symmetry presented in Section II.B, the spin-0 graviton is eliminated, so the two issues, instability and strong coupling  mentioned above,
   are absent in the gravitational sector, but the strong coupling problem re-appears  when
  coupled with matter \cite{HW11,LWWZ}. To solve the problem,  one way is to introduce a new energy scale $M_*$, as shown in Fig.\ref{fig1},  first proposed  for the non-projectable case \cite{BPSc}. Therefore, this version  is
  self-consistent. It was shown that it is also consistent with all the solar system tests \cite{LMWZ}, provided the universal coupling with matter is adopted \footnote{It should be noted that the consistency of this version of the theory with solar system 
  tests excludes the possibility of minimal coupling with matter \cite{LMWZ}.}. However, for such a coupling, the consistency with cosmological observations have not been worked out, yet, although a preliminary study indicates that a more general 
  coupling may be needed \cite{MWWZ}. An interesting problem  in this version of the theory is the physical  interpretations of the pre-Newtonian potential $\varphi$ and the U(1) gauge field $A$. Such understanding may also be able to shed lights
  on the coupling of the gravitational sector with matter. Along this direction, the non-relativistic gauge/gravity correspondence found recently in \cite{HO15} may provide some hints. 
 
  In the non-projectable case without the local U(1) symmetry presented in Section II.C, {\em the healthy extension}, it is shown that the theory is self-consistent and also consistent not only with solar system tests and cosmological observations,
  but also with the binary pulsar and gravitational wave observations \cite{YBBY,YBYB,YYP}. The strong coupling problem \cite{PS09} can be solved by introducing a new energy scale $M_*$, as shown explicitly in \cite{BPSc}, by properly choosing
  the coupling constants. One of the concerning of this version is the possibility to introduce  a hierarchy \cite{KP13}, although it was argued that this is technically quite nature \cite{BPSb,BPSc}. Another question is  the large
   number of the coupling constants (which is about 100). With such a large number, the question of the prediction power   of the theory raises again, although in the IR  only five of   them are relevant \cite{BPSa,BPSb}. 
  
  In the non-projectable case with the local U(1) symmetry presented in Section II.D, the spin-0 graviton is absent only in particular cases \cite{ZSWW,LMWZ,MSW}. But, it is always   stable in a large region of the parameters space,
   whenever it is present. The strong coupling problem in general also exists, but
  can be  solved by also introducing a new energy scale $M_*$, as one did in the last two versions. By softly breaking the detailed balance condition, the number of the coupling constants can be significantly reduced (to 15) \cite{ZSWW,ZWWS},
  while the power-counting renormalizability is still valid. It is interesting to note that this is not the case without the local U(1) symmetry, the healthy extension, as shown explicitly in \cite{ZSWW}.  It was shown that it is also consistent with all 
  the solar system tests, provided the universal coupling with matter is adopted  \cite{LMWZ}. As in the projectable case, for such a coupling, the consistency of the theory with cosmological observations has not been worked out, yet \cite{MWWZ}. 
  It is interesting to note that this version has been embedded into string theory recently by using the non-relativistic gauge/gravity correspondence \cite{JK1,JK2}. I has been also shown that it has one-to-one correspondence to the 
  dynamical Newton-Cartan  geometry   \cite{HO15}. 
  
 With all the above in mind, we have considered the recent developments of Ho\v{r}ava gravity in  three different but also related areas: (a) black holes and their thermodynamics in gravitational theories with broken Lorentz invariance, including
 Ho\v{r}ava theory; (b) non-relativistic gauge/gravity correspondence in the framework of Ho\v{r}ava gravity; and (c)  quantization of  Ho\v{r}ava theory. 
 It was somehow very surprising when it was first discovered that black holes 
 can exist (theoretically)  in gravitational  theories with broken Lorentz invariance  \cite{BS11,BJS}, as one would expect that one can explore the physics arbitrarily near the singularity, once particles with arbitrarily large speeds are allowed. Then, a natural
 question is whether such a black hole has entropy or not? Using the same arguments as we did in GR, it is not difficult to be convinced that it should have. Otherwise,  the second law of thermodynamics will be violated \cite{BBMb}. This in turn
 raises several other interesting questions, and one of them is: how the four laws of thermodynamics look like now? The first law in the neutral (without charges) case can be  generalized,  after a new definition of surface gravity given by
 Eq.(\ref{3.10}) is adopted \cite{BBMb,CLMV}. But, for charged black holes, such a generalization is still absent \cite{DWW}. 
 Recent investigations \cite{DWWZ,DL16,Cropp16} with a more general dispersion relation showed that both of the temperature 
 and surface gravity obtained by considering peering behavior of rays near the universal horizons depend on the parameter $z$, which characterizes the leading order $k^{2z}$  of the dispersion relation [cf. Eqs.(\ref{3.11}) and (\ref{3.13})]. 
 Then, even in the neutral case, it seems that the entropy $S$ of the black hole should be also $z$-dependent, if the first law, $dE = T dS$,  still holds. 
 
 For the zeroth law, it holds automatically in all the examples considered so far, as they are either in static spacetimes or in rotating (2+1)-dimensions, in which the universal horizons always occur on a  
 hypersurface on which we have $r = r_{UH} =$ constant. Kinematic considerations prove that  this is also the case  in general \cite{UH10},  although 
 concrete examples in more general spacetimes have not been found so far.  
 
 For the second law of thermodynamics, Blas and Sibiryakov  considered two possibilities where the missing entropy can be found  \cite{BS11}:
(i) It is accumulated somewhere inside the black hole (BH). BS studied the stabilities of the universal horizons and found that they are linearly stable.
 But, they argued that after nonlinear effects are taken into account, these horizons will be turned into singularities with
 finite areas. One hopes that in the full Ho\v{r}ava gravity this singularity is resolved into a high-curvature region of finite
width accessible to the instantaneous and fast high-energy modes. In this way the BH thermodynamics can be saved.
(ii) A BH has a large amount of static long hairs, which have tails that can be measured outside the horizon \cite{DTZ}. After measuring
them, an outer observer could decode the entropy that had fallen into the BH. However, BS found that spherically symmetric hairs do not exist, and to have this
scenario to work, one has to use non-spherically symmetric hairs.  Clearly, to have a better understanding of the second law, much work in this direction needs to be done.  
The same is true for the investigations of the  third law. 

The investigations of the non-relativistic gauge/gravity duality in the framework of Ho\v{r}ava gravity is still in its infancy, and various open issues  remain, including the corresponding QFTs for
the versions of Ho\v{r}ava gravity without the U(1) symmetry, their dictionaries, and so on. By flipping the logic, another important question is: can we get deeper insights about the quantization of
Ho\v{r}ava gravity from some well-known systems of QFTs through this correspondence? The answer seems very positive \cite{EH13,EH16}, but systematic investigations along this direction are still absent. 

Quantizing gravity is the main motivation for Ho\v{r}ava  to propose his theory in 2009 \cite{Horava}, but a rigorous proof of its  renormalizability is still absent, although  it is power-counting renormalizable by construction. 
So far, it has been shown that it is renormalizable only in a few particular cases. These include  (1+1)-dimensional spacetimes both with and without the projectability condition \cite{Li14,Li16,AGSW}, and spacetimes with the projectable 
condition  \cite{BBHSS}. It is still an open question how to generalize such studies to other cases, such as the one without the projectability conditions and the ones with the local U(1) symmetry. 
In addition, what is the RG flow? So far,  we have assumed that  the theory can arbitrarily approach  GR in the IR. But, without working  out the RG flow in detail, we really 
do not know if this is indeed the case or not. Currently, only very limited situations  were considered   \cite{CRS,OR09}.

Other important issues include how the theory couples with matter \cite{KP13,LMWZ}, what are the effects of the Lorentz violation \cite{PT14,BPSd,KS,Afshordi}. To the latter, we need at least  to stay below
current experimental constraints \cite{Liberati13,LZbreaking}. As mentioned in the Introduction, this is not an easy task at all \cite{Collin04,PT14,BPSd,KS,Afshordi}.  To have a proper understanding of these effects, the
coupling of  Ho\v{r}ava gravity    with matter will play a crucial role. 

Therefore, although  Ho\v{r}ava gravity seems to be a very interesting and promising alternative in quantization of gravity, for it to be really a viable theory, there is still a long list of questions that need  to be  addressed
properly.

  \section*{Acknowledgements}
  
  The author would like to express his gratitude to his collaborators in this fascinating field. They include Elcio Abdalla, Ahmad Borzou, Ronggen Cai, Gerald Cleaver, 
  Chikun Ding,  Otavio Goldoni, Jared Greenwald, A.Emir Gumrukcuoglu, Yongqing Huang, Jiliang Jing, Jonatan Lenells, Bao-Fei Li, Kai Lin, JianXin Lu, Roy Maartens, Shinji Mukohyama, Antonios Papazoglou, 
  V. H. Satheeshkumar,  M. F. da Silva,   Fu-Wen Shu, Miao Tian,  David Wands, Xinwen Wang, Qiang Wu, Yumei Wu,   Zhong-Chao Wu,  Jie Yang, Razieh Yousefi,  Wen Zhao, and Tao Zhu. He would also like to 
  thank N. Afshordi, D. Blas,  R.H. Brandenberger, B. Chen, P. Ho\v{r}ava, T. Jacobson, E.B. Kiritsis, J. Kluson,  K. Koyama,  S. Liberati, K.-i. Maeda, C. M. Melby-Thompson, A. Padilla, M. Pospelov,
  O. Pujolas, E.N. Saridakis,    M. Sasaki, S. Sibiryakov, A.M. da Silva,  J. Soda,  T.P. Sotiriou  and M. Visser, for  valuable discussions, comments and suggestions on the subjects presented here. The author
  would like  to give  his particular thanks  to   Dr. Elias C. Vagenas, the Review Editor of IJMPD, for his kind invitation  to write this review article and his patience and strong support during the whole process. 
   This work is supported in part  by Ci\^{e}ncia Sem Fronteiras, Grant No. A045/2013 CAPES, Brazil,  and Chinese NSF,  Grant Nos. 11375153 and  11173021.

   \section*{Appendix A: Lifshitz Scalar Theory}
  \renewcommand{\theequation}{A.\arabic{equation}} \setcounter{equation}{0}
     
 To introduce Lifshitz scalar theory, let us begin with a free relativistic scalar field in a (3+1)-dimensional spacetime. For the sake of simplicity and without loss of the 
 generality, we also assume that the spacetime is flat, so the action takes the form, 
 \bq
 \lb{A.1}
 S^{{\mbox{(free)}}}_{\phi} = \frac{1}{2} \int{dt d^3x \left(\dot{\phi}^2 + \phi \Delta \phi\right)},
 \eq
 where $\Delta (\equiv \partial_i^2)$ denotes the Laplacian operator. Clearly, it is  invariant under the  general transformations (\ref{2.18a}). Under the isotropic rescaling,
 \bq
 \lb{A.2}
 t \rightarrow b^{-1}t, \quad  x^i \rightarrow b^{-1}x^i, \quad \phi \rightarrow b\phi,
\eq
$S_{\phi}$ is also invariant $S_{\phi} \rightarrow S_{\phi}'$. Denoting the units of  length, time and mass, respectively,  by $L, T$ and $M$, we find that the speed $c$, energy $E$, and Planck 
constant $h$ have the dimensions, 
\bqn
\lb{A.3}
\left[c\right] &=& \frac{[\Delta x]}{[\Delta t]} = \frac{L}{T}, \quad \left[E\right] = \left[mc^2\right] = \frac{ML^2}{T^2},\nb\\
 \left[h\right] &=& \frac{[E]}{[\nu]} = \frac{ML^2T^{-2}}{T^{-1}} = \frac{ML^2}{T}.
\eqn
 Such, choosing the natural units $c = h = 1$ implies
 \bq
 \lb{A.4}
 L = T, \quad E = M, \quad
 L = M^{-1} = E^{-1}.
 \eq
 In the rest of this section, we shall use the natural units. Then, for a process with $\Delta\phi \simeq E$, we have
 \bqn
 \lb{A.5}
 \Delta t &\simeq& E^{-1}, \quad \Delta \ell \equiv \sqrt{\Delta x^2 + \Delta y^2 + \Delta z^2} \simeq E^{-1},\nb\\
 \frac{\Delta\phi}{\Delta t} &\simeq&   \frac{\Delta\phi}{\Delta x^i} \simeq \frac{\Delta\phi}{\Delta \ell} \simeq E^2.
 \eqn
 Hence, we find that
 \bqn
 \lb{A.6}
&&  \int_0^E{dt d^3x \dot\phi^2} \simeq \Delta t \left(\Delta\ell\right)^3 \left(\frac{\Delta\phi}{\Delta t}\right)^2 \nb\\
&& ~~~~~~~~~~~~~~~ \simeq E^{-1} \cdot E^{-3} \cdot \left(E^2\right)^2  \simeq {\cal{O}}(1),\nb\\
&& S^{{\mbox{(free)}}}_{\phi} = \frac{1}{2} \int{dt d^3x \left(\dot{\phi}^2 + \phi \Delta \phi\right)}  \simeq {\cal{O}}(1). ~~~~~
 \eqn
 Thus, the integral $S^{{\mbox{(free)}}}_{\phi}$ is always finite no matter how large $E$ will be, which is called {\em power-counting renormalizable}. What will happen when self-interaction is taken into 
 account? For example, let us consider the term,
 \bq
 \lb{A.7}
 S_{\phi}^{{\mbox{(s.i.)}}} = \int{dtd^3x\left(\phi^2 \Delta\phi\right)},
 \eq
which is scaling as
 \bqn
 \lb{A.8}
 S_{\phi}^{{\mbox{(s.i.)}}} &\rightarrow&  \left[dt\right] \left[dx^i\right] \left[\phi^2\right]  \left[\Delta\right] \left[\phi\right] = b^{-1-3+2+2+1}  \nb\\
 &=& b,
 \eqn
 under the rescaling (\ref{A.2}). Then, we find that
 \bqn
 \lb{A.9}
 S_{\phi}^{{\mbox{(s.i.)}}} &=& \int_0^E{dtd^3x\left(\phi^2 \Delta\phi\right)} \simeq E,
 \eqn
which becomes unbounded as $E \rightarrow \infty$, and is said (power-counting) non-renormalizable. On the other hand, as $E \rightarrow 0$, we have $S_{\phi}^{{\mbox{(s.i.)}}} \rightarrow 0$, which is said
{\em irrelevant} in low energy limit \cite{Pol}. In general, for any given operator ${\cal{O}}_{\phi}$, if $ \int_0^E{dtd^3x\; {\cal{O}}_{\phi}}$ scales as 
\bqn
 \lb{A.10}
  \int_0^E{dtd^3x\; {\cal{O}}_{\phi}} \simeq b^{\delta},
 \eqn
 under the rescaling (\ref{A.2}), we have \cite{Pol}
 \bqn
 \lb{A.11}
 &&  \int_0^E{dtd^3x\; {\cal{O}}_{\phi}} \simeq E^{\delta} \nb\\
 && ~~~ =   \begin{cases}
 \infty, &  \delta > 0,\;\;  {\mbox{Non-renormalizable}},\cr
 {\mbox{finite}}, & \delta = 0, \;\; {\mbox{Strictly renormalizable}},\cr
 0, & \delta < 0,  \;\; {\mbox{Super-renormalizable}},\cr
 \end{cases} ~~~~~~~
 \eqn
as $ E\rightarrow \infty$. On the other hand, as $E \rightarrow 0$, we find that 
 \bqn
 \lb{A.12}
 &&  \int_0^E{dtd^3x\; {\cal{O}}_{\phi}} \simeq E^{\delta} \nb\\
 && ~~~ =   \begin{cases}
 0, &  \delta > 0,\;\;  {\mbox{irrelevant}},\cr
 {\mbox{finite}}, & \delta = 0, \;\; {\mbox{marginal}},\cr
 \infty, & \delta < 0,  \;\; {\mbox{relevant}}.\cr
 \end{cases} ~~~~~~~
 \eqn
 In the rest of this section, we shall use frequently the terminology given above without any further explanations. For more details, see \cite{Pol}. In general, we have
 \bqn
 \lb{A.13}
 S_{\phi}^{{\mbox{(total)}}} &=&  S^{{\mbox{(free)}}}_{\phi} +  S_{\phi}^{{\mbox{(s.i.)}}}  + ...,\nb\\
 &=&  {\cal{O}}(1) + E + E^2 + ...  \rightarrow \infty,
 \eqn
 as $ E\rightarrow \infty$, that is, the theory is not power-counting renornalizable. 

To make the theory renormalizable, Lifshitz \cite{Lifshitz,Lifshitz2}  added the following non-relativistic term into the free action (\ref{A.1}),
\bq
\lb{A.14}
\Delta S_{\phi} \equiv \frac{1}{2} \int {dt d^3x\left(\frac{\phi\Delta^z\phi}{M_*^{2z-2}}\right)},
\eq
where $z (\ge 2)$ is an integer, and  $M_*$ is a constant, which represents the energy suppressing the  operator $\phi\Delta^z\phi$. Thus, for an integration with $\Delta \phi \simeq E$, now we have 
\bqn
\lb{A.15}
 \frac{1}{M_*^{2z-2}}\left(\frac{\phi\Delta^z\phi}{\phi\Delta\phi}\right)  &\simeq& \left(\frac{E}{M_*}\right)^{2(z-1)} \nb\\
&=&  \begin{cases}
< 1, & E < M_{*},\cr
> 1, & E > M_{*}.\cr
\end{cases}
\eqn
Therefore, when $E \ll M_*$, we have 
\bqn
\lb{A.16}
S_{\phi} &\equiv&   
 \frac{1}{2} \int_0^E{dt d^3x \left(\dot{\phi}^2 + \phi \Delta \phi + \frac{\phi\Delta^z\phi}{M_*^{2z-2}}\right)}  \nb\\
 &\simeq& \frac{1}{2} \int_0^E{dt d^3x \left(\dot{\phi}^2 + \phi \Delta \phi\right)},
 \eqn
which  is finite and invariant under the relativistic scaling (\ref{A.2}), as shown above. However, when $E \gg M_{*}$, we have 
\bq
\lb{A.17}
S_{\phi}  \simeq  \frac{1}{2} \int_0^E{dt d^3x \left(\dot{\phi}^2 + \frac{\phi\Delta^z\phi}{M_*^{2z-2}}\right)},  
\eq
which is invariant only under the anisotropic rescaling between time and space,  
\bq
\lb{A.18}
 t \rightarrow b^{-z}t, \quad  x^i \rightarrow b^{-1}x^i, \quad \phi \rightarrow b^{\alpha_{\phi}}\phi.
\eq
Under this new rescaling, we find
\bqn
\lb{A.19}
  \int_0^E{dt d^3x \dot\phi^2} \simeq b^{-z-3 + 2z + 2\alpha_{\phi}} = b ^{z-3 + 2\alpha_{\phi}},
  \eqn
where $\alpha_{\phi} \equiv \left[\phi\right]$. Thus, the scaling-invariance of this term requires 
\bq
\lb{A.20}
\alpha_{\phi} = -\frac{1}{2}(z-3).
\eq
Thus, for $z = d = 3$ the scalar field becomes dimensionless, and the interacting term (\ref{A.7}) now scales as
\bqn
\lb{A.21}
S_{\phi}^{{\mbox{(s.i.)}}} \simeq b^{-z-3 + 2\alpha_{\phi} + 2+ \alpha_{\phi}} = b^{-4},
\eqn
that is, under the new scaling, the interaction term $S_{\phi}^{{\mbox{(s.i.)}}}$ is scaling with an negative power of $b$. As a result, we have 
\bqn
\lb{A.22}
S_{\phi}^{{\mbox{(s.i.)}}} \simeq E^{-4/3} \rightarrow 0,
\eqn
as $E \rightarrow \infty$, where we assumed that $\Delta t \simeq E^{-1} \simeq p^{-1/3}$, where $\Delta \ell \simeq p^{-1}$. Thus, under the new scaling, the non-renormalizable term $S_{\phi}^{{\mbox{(s.i.)}}}$ now becomes 
renormlizable! Adding all the relevant terms into (\ref{A.14}), we find that the quadratic action
\bqn
\lb{A.23}
S_{\phi}^{{\mbox{(bare)}}}  = \frac{1}{2}\int{dt d^3x \left(\dot\phi^2 + \phi\hat{\cal{O}}_{\phi}\phi\right)},
\eqn
is power-counting renormalizable, where
\bq
\lb{A.24}
\hat{\cal{O}}_{\phi} \equiv g_3 \frac{\Delta^3}{M_{*}^4}  - g_2 \frac{\Delta^2}{M_*^2} + c^2 \Delta - m^2,
\eq
where $g_3, \; g_2, \; c $ and $m$ are coupling constants. Similarly, we can add the potential term, $V(\phi)$, into the above action, so finally we have
\bqn
\lb{A.23a}
S_{\phi}^{{\mbox{(total)}}}  = \frac{1}{2}\int{dt d^3x \left(\dot\phi^2 + \phi\hat{\cal{O}}_{\phi}\phi - V(\phi)\right)}, ~~~~
\eqn
which is clearly also power-counting renormalizable \cite{Horava,Visser,VisserA,AH07}. Note that in the most general case all the coefficients in Eq.(\ref{A.24}) should be functions of $\phi$ \cite{AH07,FIIKa,FIIKb}. 
In addition, all n-point interaction  of the forms  \cite{FIIKa,FIIKb},
\bq
\lb{A.24b}
S_{\phi}^{(z_i, n)} \equiv \lambda_{(z_i, n)}\int{dtd^3x \left(\Delta^{z_i}\phi^n\right)},
\eq
should be also included into the action where $z_i \le z$, which contains $2z_i$ spatial derivatives acting on n $\phi$'s. The dimension of $\lambda_{(z_i, n)}$ is
\bq
\lb{A.25}
\left[\lambda_{(z_i, n)}\right] = \frac{2(z-z_i)}{z},
\eq
which is always non-negative  for $z_i \le z$. As a result, this term is power-counting renormalizable. Once all such terms are included into the total action, it was shown that  there are an infinite number of UV divergent terms
in one-loop calculations \cite{FIIKa,FIIKb}, and to eliminate these divergences, one needs to introduce an infinite number of counterterms, which make the predictability of the theory questionable. To overcome this problem, 
one way is to introduce the shift symmetry \cite{FIIKa,FIIKb},
\bq
\lb{A.26}
\phi \rightarrow \phi + \phi_0,
\eq
where $\phi_0$ is a constant. Once this symmetry is imposed, it can be shown that only finite terms of the forms (\ref{A.24}) satisfy this condition. Hence, in this case only a finite number of counterterms are needed, and the theory
becomes predictable. 

Recently, the theory of Lifshitz scalar fields has been intensively studied and made applications to various cases \cite{LST3,LST4,LST5,LST6,LST7,LST8,LST9}.


\end{document}